\documentclass{emulateapj}

\usepackage{psfig}
\usepackage{natbib}
\usepackage{txfonts}
\usepackage{longtable}
\usepackage{graphicx}

\begin{document}
\def\lsun{L_{\sun}}
\def\msun{M_{\sun}}
\def\simle{{\mathop{\stackrel{\sim}{\scriptstyle <}}\nolimits}}
\def\simgr{{\mathop{\stackrel{\sim}{\scriptstyle >}}\nolimits}}
\def\lesim{{\mathop{\stackrel{\scriptstyle <}{\sim}}\nolimits}}
\shorttitle{VLA observations of NH$_3$ in IRDCs II: Kinematics}
\shortauthors{Ragan et al.}

\title{Very Large Array Observations of Ammonia in Infrared-Dark Clouds II: Internal Kinematics}

\author{Sarah E. \ Ragan$^{1,2}$, Fabian Heitsch$^3$, Edwin A. Bergin$^2$, \& David Wilner$^4$}

\affil{
$^1$ Max Planck Institute for Astronomy, K\"{o}nigstuhl 17, 69117 Heidelberg, Germany
\break email: ragan@mpia.de \\
$^2$ Department of Astronomy, University of Michigan, 830 Dennison Building, 500 Church Street, Ann Arbor, MI, 48109 USA \\
$^3$ Department of Physics and Astronomy, University of North Carolina-Chapel Hill, CB 3255 Phillips Hall, Chapel Hill, NC, 27599 USA \\
$^4$ Smithsonian Center for Astrophysics, Mail Stop 42, 60 Garden street, Cambridge, MA, 02138 USA
}

%\author{Sarah E. Ragan\altaffilmark{1,2}, Fabian Heitsch\altaffilmark{3}, Edwin A. Bergin\altaffilmark{1}, and David Wilner\altaffilmark{4}}
%\altaffiltext{1}{Department of Astronomy, University of Michigan, 830 Dennison Building, 500 Church Street, Ann Arbor, MI, 48109 USA}
%\altaffiltext{2}{Max Planck Institute for Astronomy, K\"{o}nigstuhl 17, 69117 Heidelberg, Germany \email{ragan@mpia.de}}
%\altaffiltext{3}{Department of Physics and Astronomy, University of North Carolina-Chapel Hill, CB 3255 Phillips Hall, Chapel Hill, NC, 27599 USA}
%\altaffiltext{4}{Smithsonian Center for Astrophysics, Mail Stop 42, 60 Garden street, Cambridge, MA, 02138 USA}

\begin{abstract}

Infrared-dark clouds (IRDCs) are believed to be the birthplaces of rich clusters and thus contain the earliest phases of high-mass star formation.  We use the Green Bank Telescope (GBT) and Very Large Array (VLA) maps of ammonia (NH$_3$) in six IRDCs to measure their column density and temperature structure (Paper 1), and here, we investigate the kinematic structure and energy content. We find that IRDCs overall display organized velocity fields, with only localized disruptions due to embedded star formation. The local effects seen in NH$_3$ emission are not high velocity outflows but rather moderate (few km s$^{-1}$) increases in the line width that exhibit maxima near or coincident with the mid-infrared emission tracing protostars.  These line width enhancements could be the result of infall or (hidden in NH$_3$ emission) outflow. Not only is the kinetic energy content insufficient to support the IRDCs against collapse, but also the spatial energy distribution is inconsistent with a scenario of turbulent cloud support. We conclude that the velocity signatures of the IRDCs in our sample are due to active collapse and fragmentation, in some cases augmented by local feedback from stars.
\end{abstract}

\section{Introduction}
\label{intro}

Star formation has been the focus of observational and theoretical studies for decades, but still the conditions under which this process commences are quite uncertain. The identification of objects in different evolutionary stages, such that a sequence can be constructed, is the essential observational ingredient needed to test theoretical scenarios. In the solar neighborhood, it is possible to resolve the precursors to stars (or multiple systems), known as pre-stellar cores, but the counterpart in massive regions has to date been difficult to isolate. With the recent surveys by {\em Spitzer} in the mid-infrared and advancement of millimeter and radio interferometric arrays, progress in identifying objects in various early phases of massive star formation has been rapid.

Infrared-dark clouds (IRDCs), the densest parts of molecular cloud complexes embedded within Galactic spiral arms \citep{Jackson_galdistr_IRDCs}, are believed to host these earliest stages of clustered star formation. Studies in the infrared \citep[e.g.][]{Perault1996, egan_msx, Ragan_spitzer, ButlerTan2009, Peretto2009}, millimeter continuum \citep[e.g.][]{rathborne2006,Vasyunina2009}, and molecular lines \citep[e.g.][]{carey_msx,Carey_submmIRDC, ragan_msxsurv,Pillai_ammonia,Sakai2008,DuYang2008} have shown that IRDCs contain from tens to thousands of solar masses of dense (N(H$_2$) $\sim$ 10$^{22-23}$ cm$^{-2}$) material and have the right physical conditions (T $<$ 15\,K, n $>$ 10$^5$~cm$^{-3}$) to give rise to rich star clusters, i.e. clusters which can potentially host massive (M $> 10 \msun$) stars. 

Star formation is dynamical by nature \citep[see][for a review of the important processes]{McKeeOstriker2007}, but observational tests of dynamics are complicated by the projection of this three-dimensional process onto the two-dimensional plane of the sky.  Molecular line emission -- from a number of molecules excited in the cold environments of molecular clouds -- is the key tool to help disentangle the problem along the line of sight.  Ammonia (NH$_3$) has been a particularly useful 
probe in molecular clouds \citep{HoTownes}, as it not only provides kinematic information but also serves as a cloud thermometer \citep{Walmsley1983, Maret_ammonia2009}.  Ammonia has been used widely to study local clouds \citep[e.g.][]{MyersBenson1983,Ladd1994,Wiseman1998, Jijina1999, Rosolowsky_perseus, Friesen2009} and IRDCs \citep[e.g.][]{Pillai_ammonia,Devine2011}. These studies focus on the lower metastable states, ($J,K$) =  (1,1) and (2,2), sensitive to the coldest ($<$20~K) gas without any evidence of depletion.

In \citet[][hereafter Paper 1]{Ragan2011a}, we detailed Very Large Array (VLA) observations mapping six IRDCs in the NH$_3$ ($J,K$) = (1,1) and (2,2). We used the maps to produce column density and gas temperature profiles. With between 4 and 8$''$ angular resolution, we find that ammonia traces the absorbing structure seen at 8 and 24~$\mu$m with {\em Spitzer} \citep{Ragan_spitzer}, and there is no evidence of depletion of ammonia in IRDCs. We estimated a total ammonia abundance of 8.1$\times$ 10$^{-7}$ and found that the gas temperature is roughly constant, between 8 and 13~K, across the clouds. Here, we further our analysis of these ammonia data, focusing on the velocity structure of the clouds. The high angular resolution allows us to profile the kinematics and examine their dynamical state and stability. 

\begin{table*}
\begin{center}
\caption[Target summary]{Target summary \label{tab:targets}}
\begin{tabular}{lcccccccccc}
\hline
IRDC & RA & DEC & Distance\tablenotemark{a} & $v_{lsr}$ & rms\tablenotemark{b} & beam size & $M_{IRDC}$\tablenotemark{c} & Area & $B_{cr}$\tablenotemark{d}  \\
& (J2000) & (J2000) & (kpc) & (km\, s$^{-1}$) & (mJy) & ($\arcsec \times \arcsec$) & (10$^3$ $\msun$) & (pc$^2$) & (mG) & \\
\hline
G005.85$-$0.23 & 17:59:51.4 & -24:01:10 & 3.14 & 17.2 & 2.8 & 7.7 $\times$ 6.8 & 5.5 & 0.55 & $1.63$  \\
G009.28$-$0.15 & 18:06:50.8 & -21:00:25 & 4.48 & 41.4 & 4.8 & 8.3 $\times$ 6.4 & 1.8 & 1.8 & $1.43$  \\
G009.86$-$0.04 & 18:07:35.1 & -20:26:09 & 2.36 & 18.1 & 4.3 & 8.1 $\times$ 6.3 & 2.6 & 1.3 & $0.87$  \\
G023.37$-$0.29 & 18:34:54.1 & -08:38:21 & 4.70 & 78.5 & 2.5 & 5.7 $\times$ 3.7 & 10.9 & 4.4 & $1.05$  \\
G024.05$-$0.22 & 18:35:54.4 & -07:59:51 & 4.82 & 81.4 & 4.3 & 8.2 $\times$ 7.0 & 4.0 & 1.8 & $0.99$  \\
G034.74$-$0.12 & 18:55:09.5 & +01:33:14 & 4.86 & 79.1 & 6.8 & 8.1 $\times$ 7.0 & 5.5 & 1.6 & $1.43$  \\
\hline
\end{tabular}
\tablenotetext{1}{ Assume the kinematic near distances from \citet{ragan_msxsurv}.}
\tablenotetext{2}{ of the combined data set.}
\tablenotetext{3}{ from absorption at 8~$\mu$m \citep{Ragan_spitzer}.}
\tablenotetext{4}{ Critical field strength, see \S\ref{ss:dynamical} and eq.~(\ref{e:bcrit}).}
\end{center}
\end{table*}

\section{Data \& methods}

We obtained observations of the NH$_3$ (1,1) and (2,2) inversion transitions with the Green Bank Telescope (GBT) and Very Large Array (VLA).  The observations are described in detail in Paper 1.  The single-dish and interferometer data were combined in MIRIAD (a full description of the method is found in Paper 1). A summary of the target properties sensitivity and resolution of the combined data set is given in Table~\ref{tab:targets}.  The combined data set has a velocity resolution of 0.6~km~s$^{-1}$.  In Table~\ref{tab:targets}, we also list the estimated mass and cloud area based on 8~$\mu$m extinction, which was computed with {\em Spitzer} data in \citet{Ragan_spitzer} and the critical magnetic field strength required for support, which will be discussed in \S\ref{ss:dynamical}.

At each position, the ammonia spectra were fit with a custom gaussian fitting algorithm utilizing the IDL procedure {\tt gaussfit}.  The configuration we used for the VLA backend did not fit the entire NH$_3$ (1,1) hyperfine signature (spanning $\sim$3.6~MHz) in the bandpass (3.125~MHz). Our line-fitting routine takes a ``first guess'' line center velocity of the central line \citep[from][see Table~\ref{tab:targets}]{ragan_msxsurv} which is offset by approximately 7.7 km~s$^{-1}$ from the neighboring hyperfine components to either side. We fit each of the components independently. For the NH$_3$(2,2) lines, a single gaussian was fit to the line independently of the results of the (1,1) fit.  From these fits, we extract the peak intensity, line-center velocity, and Gaussian width of the lines at each position.

\begin{figure*}
\begin{center}
\includegraphics[scale=0.8,angle=270]{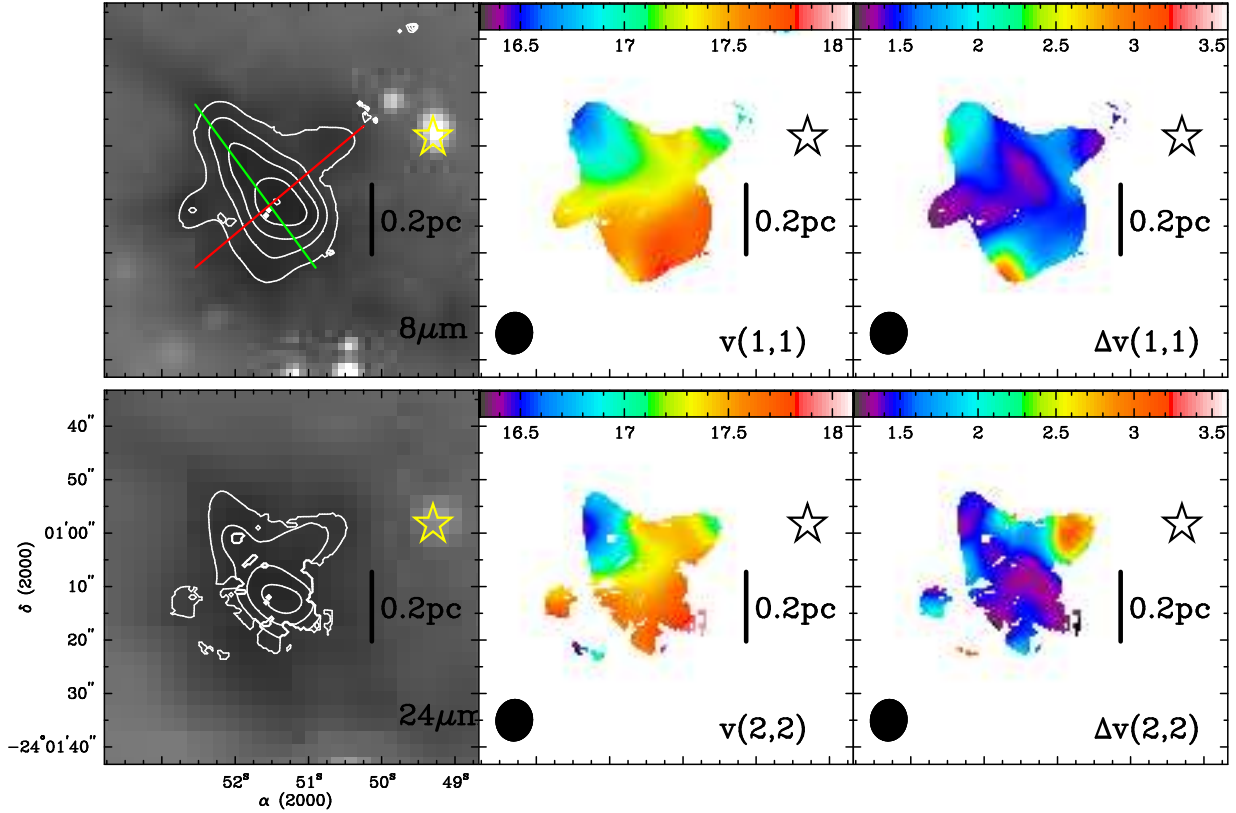}
\end{center}
\caption{\label{fig:g0585_3plot} Spectral moments in G005.85$-$0.23.  Top left: {\em Spitzer}/IRAC 8$\mu$m image with NH$_3$(1,1) integrated intensity contours overlaid. Contours begin at 0.2 Jy beam$^{-1}$ km s$^{-1}$ and increase in 0.1 Jy beam$^{-1}$ km s$^{-1}$ steps. Bottom left: {\em Spitzer}/MIPS 24$\mu$m image with NH$_3$(2,2) integrated intensity contours overlaid. Contours begin at 0.02 Jy beam$^{-1}$ km s$^{-1}$ and increase in 0.01 Jy beam$^{-1}$ km s$^{-1}$ steps. Top center: NH$_3$(1,1) centroid velocity map in km s$^{-1}$. Top right: FWHM of NH$_3$(1,1) central line in km s$^{-1}$. Bottom center: NH$_3$(2,2) centroid velocity map in km s$^{-1}$. Bottom right: FWHM of NH$_3$(2,2) line in km s$^{-1}$. The red line represents the major axis and the green line represent the minor axis. The star symbol represents a point source which only appears at 24~$\mu$m. The VLA beam is shown at the lower-left corner of the first and second moment panels.}
%The symbols in Figures 1 through 6 represent the locations of YSOs: Class I (blue square), Class II (green diamond), embedded objects (magenta diamonds), and unclassified point sources appearing only at 24~$\mu$m (star symbols). The VLA beam is represented in the lower-left corner of the first and second moment panels. }
\end{figure*}

\begin{figure*}
\begin{center}
\includegraphics[scale=0.8, angle=270]{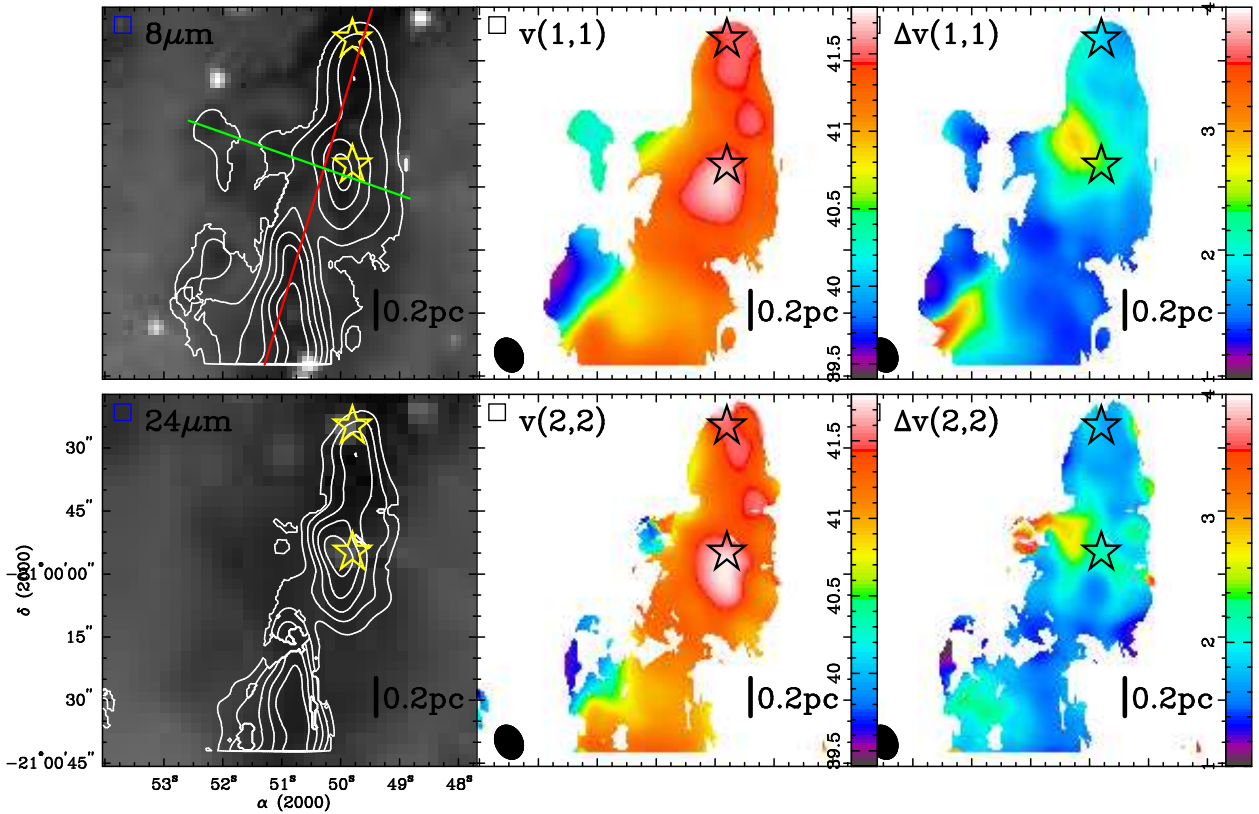}
\end{center}
\caption{Same as Figure~\ref{fig:g0585_3plot} but for G009.28$-$0.15. \label{fig:g0928_3plot}}
\end{figure*}

\begin{figure*}
\begin{center}
\includegraphics[scale=0.8,angle=270]{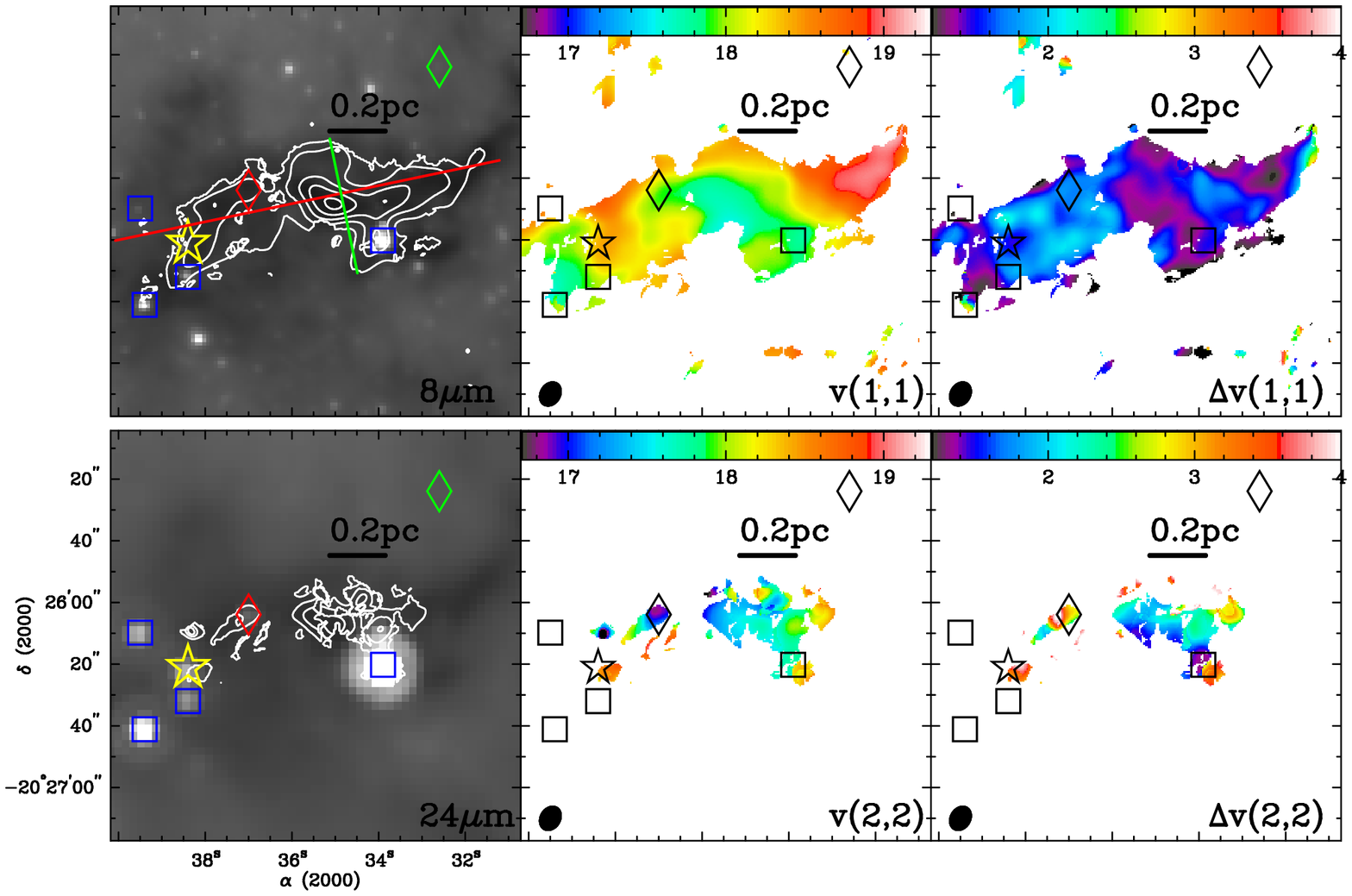}
\end{center}
\caption{Same as Figure~\ref{fig:g0585_3plot} but for G009.86$-$0.04.  NH$_3$(2,2) integrated intensity contours (overplotted in bottom-left panel) begin at 0.03 Jy beam$^{-1}$ km s$^{-1}$ and increase in 0.01 Jy beam$^{-1}$ km s$^{-1}$ steps. \label{fig:g0986_3plot}}
\end{figure*}

\begin{figure*}
\begin{center}
\includegraphics[scale=0.8,angle=270]{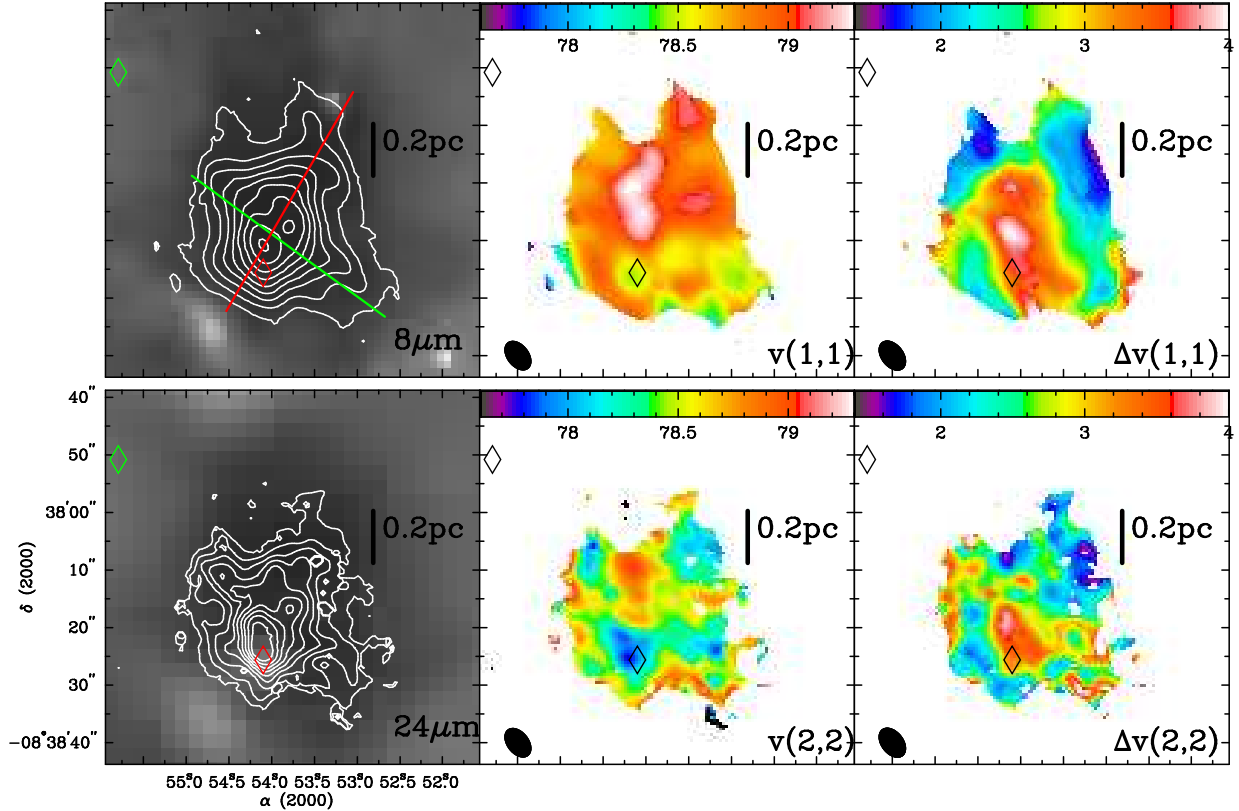}
\end{center}
\caption{Same as Figure~\ref{fig:g0585_3plot} but for G023.37$-$0.29.  \label{fig:g2337_3plot}}
\end{figure*}

\begin{figure*}
\begin{center}
\includegraphics[scale=0.8,angle=270]{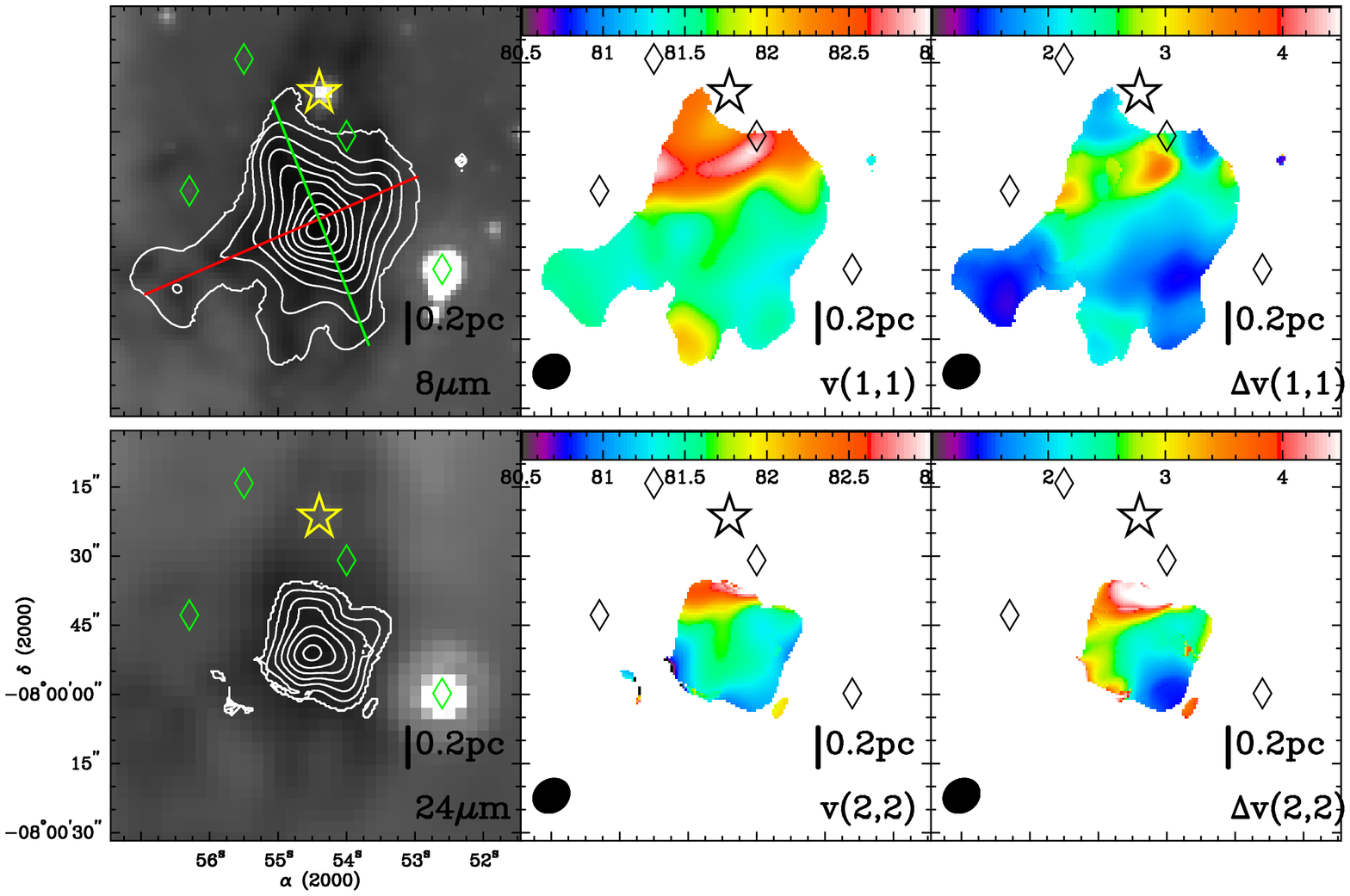}
\end{center}
\caption{Same as Figure~\ref{fig:g0585_3plot} but for G024.05$-$0.22. NH$_3$(2,2) integrated intensity contours (overplotted in bottom-left panel) begin at 0.04 Jy beam$^{-1}$ km s$^{-1}$ and increase in 0.01 Jy beam$^{-1}$ km s$^{-1}$ steps. \label{fig:g2405_3plot}}
\end{figure*}

\begin{figure*}
\begin{center}
\includegraphics[scale=0.8,angle=270]{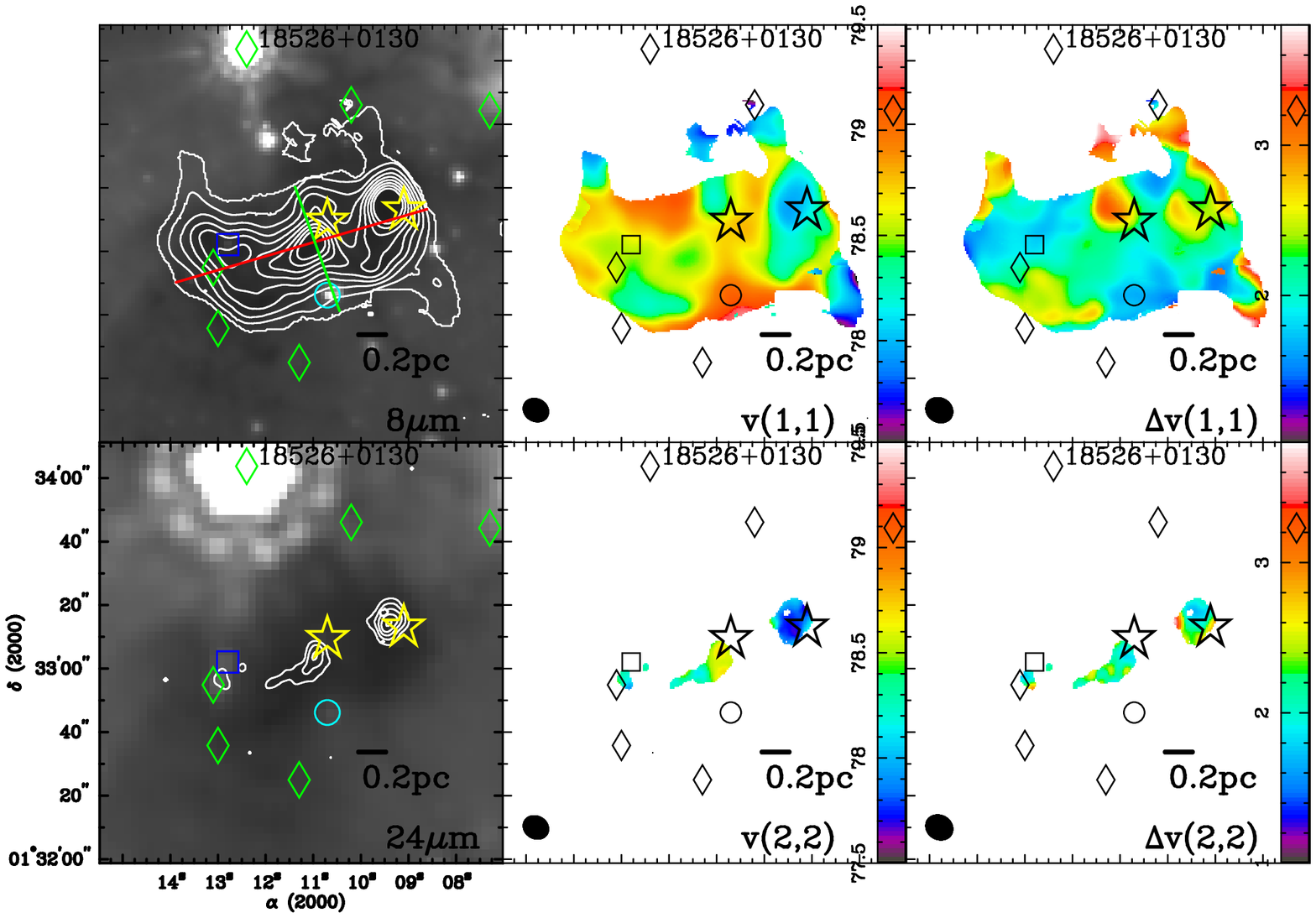}
\end{center}
\caption{Same as Figure~\ref{fig:g0585_3plot} but for G034.74$-$0.12. NH$_3$(2,2) integrated intensity contours (overplotted in bottom-left panel) begin at 0.04 Jy beam$^{-1}$ km s$^{-1}$ and increase in 0.01 Jy beam$^{-1}$ km s$^{-1}$ steps. \label{fig:g3474_3plot}}
\end{figure*}

\section{Results}

Figures \ref{fig:g0585_3plot} through \ref{fig:g3474_3plot} show the NH$_3$(1,1) integrated intensity\footnote{Since our VLA bandpass included only the central 3.125~MHz of the $\sim$3.5~MHz hyperfine signature, we observe only the central three of the five main components of the NH$_3$ (1,1) hyperfine signature. To compute the integrated intensity of the line, we assume that the missing outer-most lines are 0.22 the strength of the main lines and that the linewidths of all components are equal.} and (2,2) integrated intensity plotted over the 8 and 24~$\mu$m {\em Spitzer} images of the regions \citep[from ][]{Ragan_spitzer}, respectively, and maps of the line center velocity (first moment) and line width (second moment) for the central component of the NH$_3$ (1,1) signature. We also plot the physical scale assuming the distances in Table~\ref{tab:targets}. 

\begin{table*} 
\small
\caption{Summary of NH$_3$ (1,1) and (2,2) peak characteristics  \label{tab:velocity}}
%\centering
\begin{tabular}{lrccccccccl}
\hline
& & \multicolumn{4}{c}{Peak $\int T dv$ Position of NH$_3$ (1,1) transition} & & \multicolumn{2}{c}{NH$_3$ (2,2) line} & \\
\cline{3-6}  \cline{8-9} \\
IRDC & &  $\alpha$ & $\delta$ & v$_{lsr}$ & $\Delta$v & $\tau_m$(1,1)  & v$_{lsr}$ & $\Delta$v &  $R_p$ & Notes  \\
name & & (J2000) & (J2000) & (km s$^{-1}$) & (km s$^{-1}$) & &  (km s$^{-1}$) &  (km s$^{-1}$)   & & \\
\hline
G005.85$-$0.23 & &
17:59:51.4 &
$-$24:01:10 &
17.36$\pm$0.02 &1.41$\pm$0.03 & 4.9 &
17.49$\pm$0.04 & 1.37$\pm$0.04 & 0.1 & smooth $v$-grad.
\\
G009.28$-$0.15 & P1 &
18:06:50.8 &
$-$21:00:25 &
40.99$\pm$0.02 & 1.45$\pm$0.02 & 4.0  &
40.99$\pm$0.03 & 1.53$\pm$0.03 & 0.1 & main peak
\\
& P2 & % nw secondary peak, with 24um star
18:06:49.9 &
$-$20:59:57 &
41.70$\pm$0.07 & 2.37$\pm$0.07 & 4.0 &
41.75$\pm$0.08 & 2.14$\pm$0.08 & 0.04 & 24~$\mu$m source
\\
& P3 & % top peak, starless
18:06:49.8 &
$-$20:59:34 &
41.47$\pm$0.03 & 1.85$\pm$0.03 & 3.6 &
41.38$\pm$0.04 & 1.67$\pm$0.05  & 0.06 & 24~$\mu$m source
\\
G009.86$-$0.04 & & 
18:07:35.1 &
$-$20:26:09 &
17.75$\pm$0.04 & 1.35$\pm$0.04 & 3.3 &
17.50$\pm$0.06 & 1.64$\pm$0.07 & 0.1 & ``quiescent'' peak, 2 v-grad.
\\
G023.37$-$0.29 & &
18:34:54.1 &
$-$08:38:21 &
78.81$\pm$0.20 & 3.89$\pm$0.22 & \nodata\tablenotemark{a} &
78.16$\pm$0.15 & 3.66$\pm$0.19 & & 24~$\mu$m source
\\
G024.05$-$0.22 & &
18:35:54.4 & 
$-$07:59:51 &
81.65$\pm$0.03 & 1.96$\pm$0.03 & 2.6 &
81.62$\pm$0.05 & 2.14$\pm$0.07 & 0.05 & N-S v-grad.
\\
G034.74$-$0.12 & P1 & % nw peak (stronger)
18:55:09.5 & 
$+$01:33:14 &
77.95$\pm$0.03 & 2.46$\pm$0.03 &6.1&
77.72$\pm$0.05 & 2.08$\pm$0.05 & 0.04 & 24~$\mu$m source
\\
& P2 &
18:55:11.0 & 
$+$01:33:02 &
78.69$\pm$0.03 & 2.09$\pm$0.03 & 2.9 &
78.58$\pm$0.07 & 2.04$\pm$0.07 &  0.05 & 24~$\mu$m source
\\
\hline
\end{tabular}
\tablenotetext{1}{line saturated.}
\end{table*}
\normalsize

\subsection{Properties of individual sources}

Paper 1 demonstrates that the gas in these IRDCs exhibit uniform temperatures, changing by only a few Kelvin in a given object, not significantly more than the error.  In contrast, their velocity fields -- both the line-center velocities and line width measurements -- show a connection between the presence of embedded star formation activity and complex kinematic signatures.  The upper panels of Figures 1--6 show the (zeroth, first, and second, from left to right) moment maps derived from the NH$_3$(1,1) observations for each IRDC in the sample with symbols indicating the locations of the 24~$\mu$m point sources and other young stars, and the lower panels show the same for the NH$_3$(2,2) emission.  Table~\ref{tab:velocity} summarizes the kinematic properties of the emission peaks. Although we list only the velocity properties of the central component of the NH$_3$(1,1) line, the velocity structure traced by the satellite lines closely follows the trends seen in the central line. We also list the main line optical depth of the NH$_3$(1,1) transition, $\tau_m$(1,1), the ratio of thermal to non-thermal contributions to the pressure ($R_p$), which will be discussed in Section~\ref{discussion}, and notes about the velocity trend in the cloud or particular characteristics of the integrated intensity peak.  In this section, we discuss the centroid velocity and linewidth trends in each IRDC individually and connect the detected 24~$\mu$m point sources to the kinematic signatures.\\

For the sake of our modeling, we categorize each IRDC based on its NH$_3$(1,1) emission morphology as either a ``sphere'' for objects with an aspect ratio, $r$, close to one or a ``filament'' for objects with $r$ much greater than one. The axes used to make this distinction are indicated in Figures 1--6, and the morphological type is listed in Table~\ref{tab:fits}. For spheres, the aspect ratio is no greater than 1.1, and the elongated structures, or ``filaments,'' range from 1.6 to 2.9 in $r$.  

\noindent{\em G005.85$-$0.23:}
This source appears approximately round  ($r\sim$1.1) in the NH$_3$ (1,1) and (2,2) integrated intensity map. The peak at $\alpha (2000) = 17^{h} 59^{m} 51.4^{s}$, $\delta (2000) = -24^{\circ} 01 \arcmin 10 \arcsec$ corresponds to the position of the peak in 8~$\mu$m optical depth.  There are no 24~$\mu$m sources in the mapped region. 

The smooth gradient in centroid velocity in this IRDC permits us to straightforwardly quantify and distinguish the large-scale ordered motions and the remaining residual motion on small scales.  We show in the central panels of Figure~\ref{fig:g0585_3plot} a clear velocity gradient oriented 30 degrees east of north.  The total gradient in the NH$_3$ (1,1) emission is 1.2 km~s$^{-1}$ over 35 arcseconds, or 0.5~pc, resulting in a velocity gradient of 2.4~km~s$^{-1}$~pc$^{-1}$.  If this linear gradient is subtracted, the residual values do no exceed 0.2~km~s$^{-1}$, indicating the bulk motion dominates the dynamics of the cloud. The overall linewidth measured across the cloud is very low, between 1.3 and 1.8~km~s$^{-1}$, but it increases sharply at the edges (to $\sim$3~km~s$^{-1}$) where the centroid velocity also falls off quickly.\\

\noindent{\em G009.28$-$0.15:}
In this ``filament'' ($r\sim$1.6), there are three integrated intensity maxima: the central peak ($\alpha (2000) = 18^{h} 06^{m} 49.9^{s}$, $\delta (2000) = -20^{\circ} 59 \arcmin 57 \arcsec$, P2 in Table~\ref{tab:velocity}), which has a 24~$\mu$m source associated with it, P3 to the north (offset 25 $\arcsec$), and the maximum (P1) to the south (offset 30 $\arcsec$).  P2 is near the linewidth maximum (3.3~km~s$^{-1}$), and is also red-shifted in centroid velocity.  P1, while the strongest in integrated intensity has the lowest linewidths detected in this object (1.4~km~s$^{-1}$), and there is no associated 24~$\mu$m source. The northern integrated intensity peak is 10 $\arcsec$ away from a 24~$\mu$m point source, but the kinematic structure is not altered by its presence.  

Apart from P2 and P3, the bulk of the cloud resides at a narrow range of line center velocities, between 41 and 41.5 km s$^{-1}$. The sharpest changes in centroid velocity are located at the eastern edge of the cloud, where the line is blue-shifted by 1-1.5~km~s$^{-1}$ with respect to the bulk of the cloud at the southeast edge in both NH$_3$ (1,1) and (2,2) emission. The linewidths are also enhanced at this edge, though no YSOs are detected in this region.\\

\noindent{\em G009.86$-$0.04:}
We approximate this source as a filament, the most elongated structure ($r\sim$2.9) in our sample. The integrated intensity peak ($\alpha (2000) = 18^{h} 07^{m} 35.0^{s}$, $\delta (2000) = -20^{\circ} 26 \arcmin 09 \arcsec$) is dark at both 8 and 24~$\mu$m and corresponds to where the centroid velocity and linewidth is the lowest, all evidence for a quiescent region.  The NH$_3$ (1,1) centroid velocity field in this object is organized into two gradients in either direction from the central peak in integrated intensity. The eastern (left-hand) gradient is of magnitude $\sim$1~km~s$^{-1}$ oriented 80$^{\circ}$ east of north, and the western (right-hand) gradient is of magnitude $\sim$1.5~km~s$^{-1}$ oriented 65$^{\circ}$ west of north.  The central ``hinge'' position is indistinct in the linewidth measurement. 

Overall the velocity field in the filament is smooth, with several YSOs coincident with the ammonia emission: five east of the intensity peak, and one to the southwest. The NH$_3$ (1,1) linewidths are enhanced ($> 2$ km s$^{-1}$) in the east.  Curiously, the NH$_3$ (2,2) emission, which appears to follow the locations of the YSOs in the eastern region, exhibits an overall shift to lower line-center velocities (by 0.6~km s$^{-1}$) and higher linewidths (by 0.5~km s$^{-1}$). The optical depth of the NH$_3$(1,1) main line is below $\sim$3 in the eastern region, so it is unlikely that optical depth effects are the cause of the increased linewidth. This object appears to be undergoing cluster formation, though the part of the cloud associated with NH$_3$ (1,1) emission peak remains quiescent. 
\\

\noindent{\em G023.37$-$0.29:}
The integrated intensity map of this round IRDC ($r\sim$1.1) peaks at $\alpha (2000) = 18^{h} 34^{m} 54.1^{s}$, $\delta (2000) = -8^{\circ} 38 \arcmin 21 \arcsec$, although throughout this cloud, the lines are saturated and/or optically thick, making it impossible to derive reliable optical depths or very accurate line properties.  There is a 24~$\mu$m point source (not present at 8~$\mu$m) near the center of the region, slightly offset from the intensity peak, which is likely a deeply embedded protostar. The central region has very high linewidths (highest in the sample, 4~km~s$^{-1}$, see Figure~\ref{fig:g2337_3plot}), although because of the high optical depth of the NH$_3$ lines, these should be taken cautiously.\\

\noindent{\em G024.05$-$0.22:}
This approximately round source ($r\sim$1.1) appears centrally-peaked in line intensity (left panels of Figure~\ref{fig:g2405_3plot}) at $\alpha (2000) = 18^{h} 35^{m} 54.1^{s}$, $\delta (2000) = -7^{\circ} 59 \arcmin 51 \arcsec$, which corresponds also to the peak in 8~$\mu$m optical depth.
The NH$_3$ (1,1) maps shows a velocity gradient starting at an east-west aligned ``ridge'' slightly offset to the north from the peak of integrated intensity.  This ``ridge'' in centroid velocity also corresponds with enhanced linewidths ($\sim$3.6~km~s$^{-1}$) in both the (1,1) and (2,2) lines, though with no distinction in integrated intensity similar to what we see in G009.86$-$0.04. From the center  of the cloud across this ridge, the velocity changes by $\sim$1.1~km~s$^{-1}$, and corresponds to a gradient of 2.1 km s$^{-1}$ pc$^{-1}$. At the southern tip of (1,1) emission, there appears to be a clump with distinct red-shifted velocity but this is not detected in (2,2) emission. While there is one Class II source coincident with northern ridge, this IRDC lacks 24~$\mu$m detections (an indicator of an embedded source) anywhere in the cloud.\\

\noindent{\em G034.74$-$0.12:}
The overall velocity field of this filament ($r\sim$2.0) appears quite disorganized, particularly in locations where 24~$\mu$m point sources are detected. There are two integrated intensity peaks in this IRDC, both in the vicinity of 24~$\mu$m sources (star symbols in Figure~\ref{fig:g3474_3plot}), and both positions show high linewidth.  The strong NH$_3$ (1,1) peak in the northwest portion of the cloud ($\alpha (2000) = 18^{h} 55^{m} 09.5^{s}$, $\delta (2000) = +1^{\circ} 33 \arcmin 14 \arcsec$, P1 in Table~\ref{tab:velocity}) is directly coincident with a 24~$\mu$m point source, enhanced linewidths and a slightly blue-shifted centroid velocity.  The optical depth of the NH$_3$(1,1) main line is 6.1.  The centrally located peak ($\alpha (2000) = 18^{h} 55^{m} 11.0^{s}$, $\delta (2000) = +1^{\circ} 33 \arcmin 02 \arcsec$, P2 in Table~\ref{tab:velocity}) is offset 10 $\arcsec$ from the position of the 24~$\mu$m source and offset 15 $\arcsec$ from the nearby peak in linewidth.  The optical depth of the NH$_3$(1,1) line here is 2.9.  These two locations are also where most of the appreciable NH$_3$ (2,2) emission is found.  

\begin{figure}
\begin{center}
\includegraphics[scale=0.5]{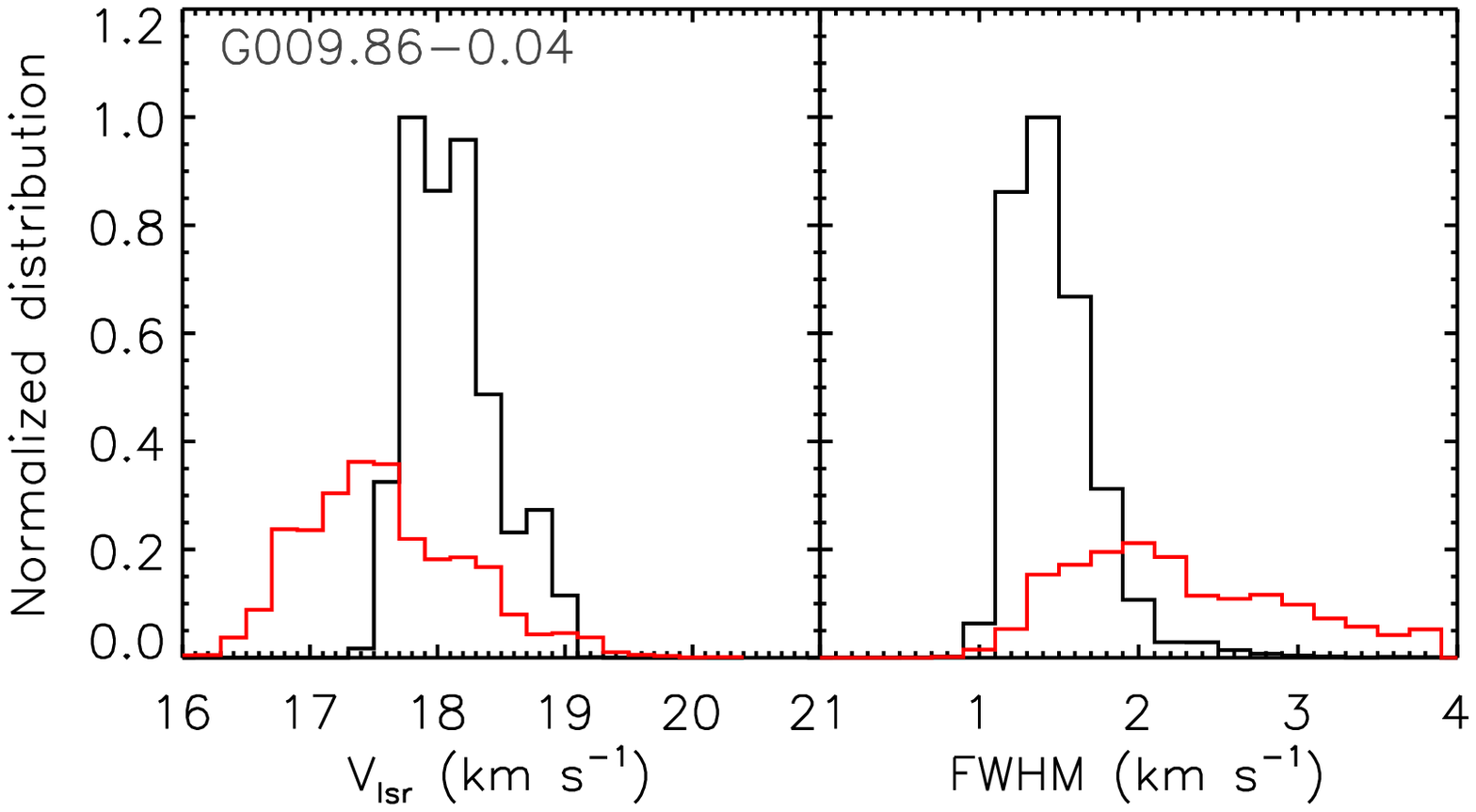}
\vspace{-1cm}
\includegraphics[scale=0.5]{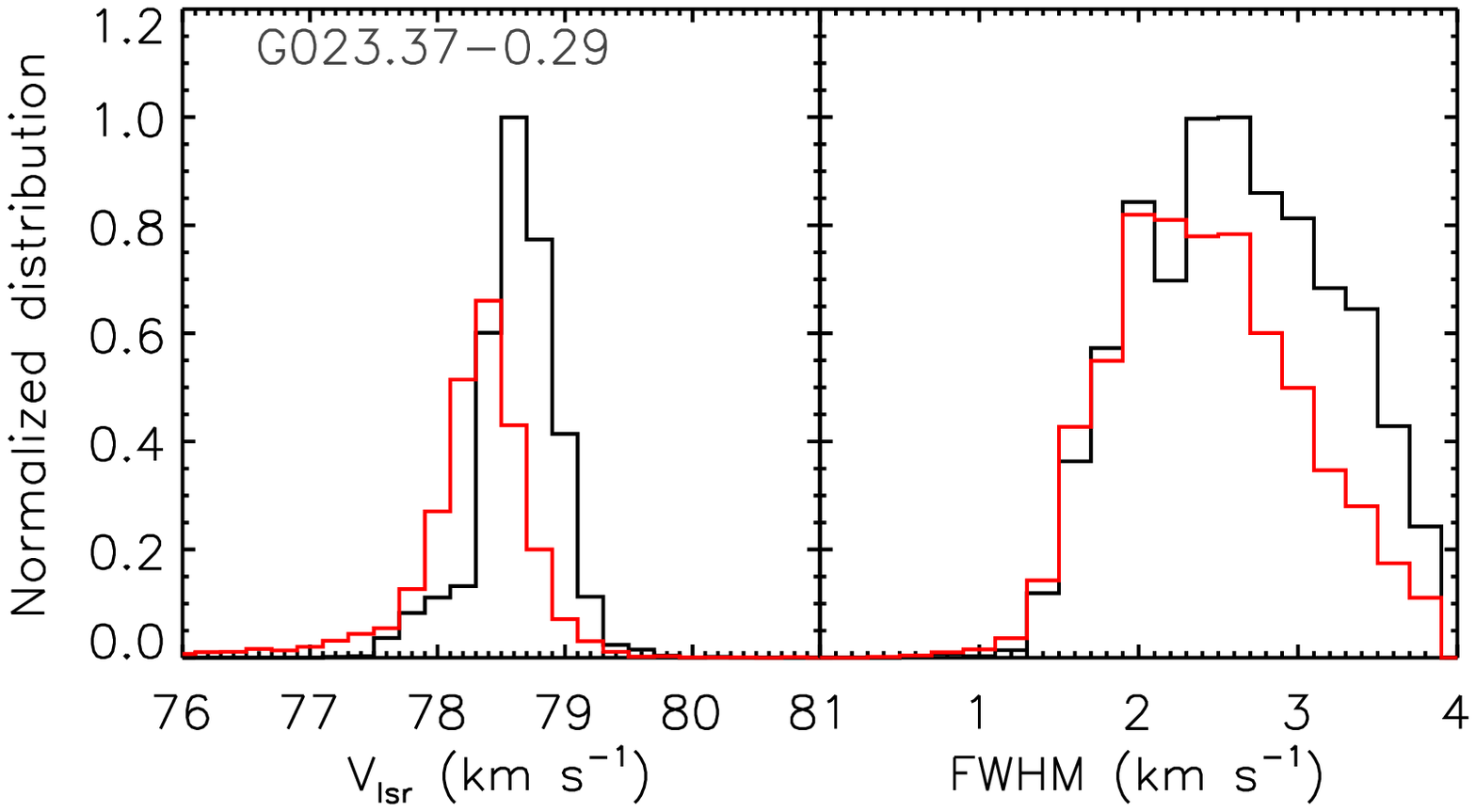}
\end{center}
\caption{Distribution of velocity fit parameters for IRDC G009.86$-$0.04 and G023.37$-$0.29 with the NH$_3$(1,1) plotted in the black histogram and the NH$_3$(2,2) distribution plotted in the red histogram. Both histograms are normalized to the number of positions with a (1,1) measurement. The left panel shows the line center velocities and the bottom panel shows the FWHM, both in km s$^{-1}$. \label{fig:g0986_vdistrib}}
\end{figure}

\begin{figure}
\begin{center}
\includegraphics[scale=0.6]{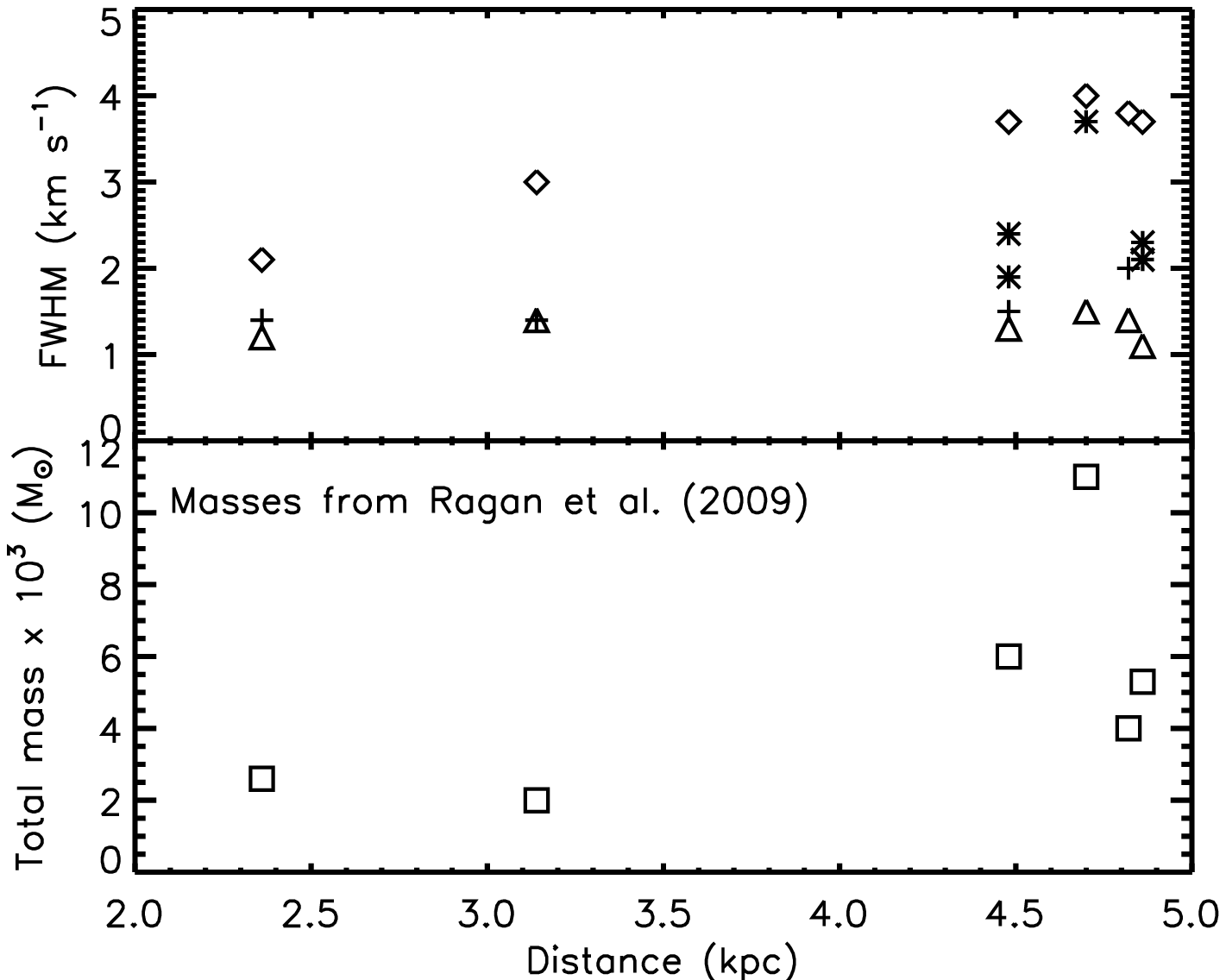}
\end{center}
\caption{Top: Linewidth vs. IRDC distance.  The linewidth at the integrated intensity peaks (marked with asterisks where there 24~$\mu$m point sources present and $+$ signs for those without 24~$\mu$m point sources) and the maximum detected linewidth overall in the cloud (marked with diamonds) both increase with increasing distance. Bottom: Total mass \citep[from 8~$\mu$m absorption,][]{Ragan_spitzer} vs. distance.}
\label{fig:dvdist}
\end{figure}

\subsection{Comparison between (1,1) and (2,2) kinematics}

As is shown in the left panels of Figures~\ref{fig:g0585_3plot} through \ref{fig:g3474_3plot} the NH$_3$ (1,1) emission tends to be more widespread than the (2,2) emission. In this section, we compare the velocity fields of the two states.  The central panels show the range in line center velocities, which generally encompass the same range, and the right panels show similar trends in linewidth. Table~\ref{tab:velocity} shows the velocity properties of the (1,1) and (2,2) at the locations of the NH$_3$(1,1) intensity peaks. 

In G009.86$-$0.04 and, to a lesser extent in G023.37$-$0.29, we find that the median NH$_3$(2,2) line center velocity is offset to lower velocities by $\sim$0.6~km s$^{-1}$ compared to the NH$_3$(1,1), and the median FWHM of the (2,2) line is higher than that of the (1,1) line by $\sim$0.8~km s$^{-1}$. Figure~\ref{fig:g0986_vdistrib} shows the distributions of line center velocity and linewidth from both the (1,1) and (2,2) maps. G009.86$-$0.04 has several young stars and 24~$\mu$m point sources in the eastern part of the cloud, which corresponds to the locations of the high linewidths and blue-shifted line center velocities in the NH$_3$(2,2) map. The optical depth of the NH$_3$(1,1) main line is less than 3 in this region, lower than typical values of 4 or 5 throughout the sample, so it is unlikely that optical depth effects are the cause of the enhanced linewidth. It may be the case that the young stars are having a dynamical effect on the slightly warmer gas probed by the NH$_3$(2,2) emission.  This does not appear to be the case in G023.37$-$0.29, where there is only a singular 24~$\mu$m point source and the lines are optically thick, which would likely limit our ability to probe near any embedded source(s).

IRDC G034.74$-$0.12 also exhibits evidence that a cluster is actively forming with the two 24~$\mu$m point sources corresponding to very strong NH$_3$ emission and several other young stars in the vicinity.  However, in these two locations of 24~$\mu$m point sources, which is also where most of the NH$_3$(2,2) emission is detected, the linewidth appears more enhanced in the NH$_3$(1,1) rather than (2,2). We note that the high optical depth of NH$_3$(1,1) at P1 (6.1) may contribute to the broadened line here. The observations of this object were the noisiest of the sample, and it is also the most distant IRDC, so we may not be sensitive to the effect we see in G009.86$-$0.04 (the nearest IRDC).

\subsection{Linewidth}
The NH$_3$ (1,1) linewidths in our sample of IRDCs are between 1.1 and 4 km s$^{-1}$, occupying the high tail of the linewidth distribution presented in the \citet{Jijina1999} survey of 264 dense cores, but on par with other ammonia studies of IRDCs: \citet{Pillai_ammonia}, who found a slightly lower range in their single-dish study, and \citet{Wang_ammonia}, who found that different cores within an IRDC exhibited different linewidths in the high-resolution study: higher linewidths near locations of embedded star formation activity and lower linewidths in quiescent regions of the cloud. For our sources, the enhanced linewidth appears to correspond to the locations of 24~$\mu$m point sources.  

We find that the linewidth increases with increasing distance to IRDCs, as was noted by \citet{Pillai_ammonia}.  In Figure~\ref{fig:dvdist}, we show that the maximum linewidth detected in IRDCs varies directly with distance, which may be a result of clumping within the beam increasing with distance.  Certainly, with {\em Spitzer} we do see objects on the 2 - 3\arcsec scale which would not be resolved with the beam (sometimes 6 - 8\arcsec in low-elevation sources).  At the same time, the minimum linewidth does not show any trend with distance.  As noted above, the integrated intensity peaks with 24~$\mu$m point sources have a higher linewidth than those peaks without, but the apparently starless peaks typically have very low linewidths, independent of distance ($\Delta$v $\sim$ 1.4~km~s$^{-1}$). Therefore, in the following discussion, we proceed with our analysis assuming that the trends are intrinsic features not determined by the distance. As shown in the lower panel of Figure~\ref{fig:dvdist}, the cloud masses increase slightly with distance, which could simply be a selection effect. 
Higher cloud masses suggest deeper gravitational potentials, which in turn would give rise to larger linewidths. Yet the total kinetic energy calculated from the averaged line widths and the centroid velocities does not change perceptibly with distance.

The dynamical properties of all IRDCs are summarized in Figure~\ref{fig:vhist}. The centroid velocity distributions (left column) are asymmetric, with tails to lower or higher velocities than the systemic velocity. This could indicate a substantial elongation along the line-of-sight, or highly asymmetric infall. The relative motions to the systemic velocity are mostly supersonic. The non-thermal velocity dispersion
\begin{equation}
  \sigma_{NT} = \sqrt{\frac{\Delta v^2}{8\ln 2}-\frac{k_BT}{\mu m_H}}
  \label{e:sigmant}
\end{equation}
is larger than the sound speed (see Paper 1 for temperatures, ranging from 8 to 13~K) for all IRDCs -- in fact, none of the velocity dispersions are
smaller than Mach 2. The distributions are peaked, with tails to high Mach numbers. For IRDCs containing sources (see  discussion above), these tails, or ``excesses'' are correlated with locations in the vicinity of the sources.

\section{Discussion}
\label{discussion}

Judging from Figures \ref{fig:g0585_3plot} through \ref{fig:g3474_3plot}, it is unlikely that one model of IRDC kinematics can fully account for the range of behaviors observed. The broad characteristics -- the overall linewidths and ranges of centroid velocities, see Table~\ref{tab:velocity} -- are roughly consistent among the clouds in the sample and with previous molecular line studies \citep[e.g.][]{Pillai_ammonia,ragan_msxsurv}. However, our sample exhibits both global trends (e.g. smooth velocity gradients) and localized effects (e.g. signatures of feedback from young embedded protostars) that can now be investigated with high angular resolution observations.

As a first approximation of stability, we compute the virial mass of the clouds ($M_{Virial} = 5 R V_{rms}^2/(3G)$), where $R$ is the radius of the cloud, $G$ is the gravitational constant, and $V_{rms}$ = 3$^{\frac{1}{2}} \Delta$V/2.35 where $\Delta$V is the average linewidth of the cloud. Virial masses for the whole clouds are typically 10$^{2-3} \msun$, and the virial parameters, $\alpha = M_{Virial}/M$, from 0.1 to 0.7, suggesting that IRDCs are bound structures prone to collapse. The cloud masses $M$ are taken from the dust extinction maps by \citet{Ragan_spitzer}.  

The linewidths observed in our sample (1.1 - 4.0 km~s$^{-1}$) are in excess of the thermal linewidth, $\sim$0.18~km~s$^{-1}$. The ratio of thermal to non-thermal pressure in the cloud, as expressed by \citet{lada_b68}, is $R_p\equiv {c_s^2}/{\sigma^2_{NT}}$, where $c_s$ is the isothermal sound speed, and $\sigma_{NT}$ is the three-dimensional non-thermal velocity dispersion.  We calculate values for $R_p$ at each of the intensity peaks in the IRDCs (see Table~\ref{tab:velocity}) and find an average value of 0.07, indicating that non-thermal pressure is dominant. ``Non-thermal effects'' encompass many things, such as infall, outflows, or systematic cloud motions (i.e. rotation) and possible ``support'' from turbulent motions \citep{AronsMax2001} or magnetic fields \citep{MousSpitzer1976}. In the following sections, we explore these effects individually, so to determine the dominant processes.

\subsection{Connecting star formation and IRDC kinematics}\label{sf_kinematics}

The IRDCs in our sample span a range of stages of star formation, some devoid of embedded sources and some hosting several. In Paper 1, we showed that the presence of young stars does not significantly affect the kinetic temperature of the gas traced by ammonia, at least beyond the $\sim$1\,K errors we estimate\footnote{We noted in Paper 1 that these observations of the (1,1) and (2,2) transitions of ammonia are not necessarily sensitive to hotter gas on the smallest scales, which (for example) may arise from a young embedded protostar heating a compact core. Gas in warm, compact regions are better probed with higher-J transitions of ammonia or other molecules.}. However, as we showed in Section 3.1, the presence of young stars, particularly 24~$\mu$m point sources, has a localized effect on the IRDC dynamics, namely an increase in linewidth at the positions of young stars and (sometimes) distinct centroid velocity components. We see no strong trend for the positions of the 24~$\mu$m point sources to have exceptionally high optical depths in NH$_3$(1,1), thus we do not expect optical depth effects to contribute strongly to linewidth enhancements.  Furthermore, starless peaks in NH$_3$ intensity (existing in all clouds except G023.37$-$0.29 an G034.74$-$0.12) tend to have the lowest linewidths.

Broadened linewidths are an indicator of increased internal motions which accompany the onset of star formation \citep{Beuther2005}. In each IRDC (except G005.85$-$0.23 and G024.05$-$0.22) of our sample, we see that the sites of 24~$\mu$m point sources are accompanied by an enhancement in NH$_3$(1,1) linewidth (in G009.86$-$0.04, the enhancement is rather seen in the NH$_3$(2,2) linewidth).  \citet{Beuther2005} found that in a molecular core with an infrared counterpart, the N$_2$H$^+$(1-0) emission \citep[known to trace dense gas similarly to NH$_3$, e.g.][]{Johnstone2010} exhibited broad line emission, whereas linewidths in infrared-dark cores -- presumably at an earlier evolutionary stage -- were significantly narrower ($\Delta$v $\sim$ 1~km~s$^{-1}$).  The increased internal motions giving rise to broadened lines can be attributed to ordered motion, such as infall, outflow or rotation.

\begin{table*} 
\caption{Summary of fit results \label{tab:fits}}
\begin{center}
\begin{tabular}{lrccccc}
\hline
IRDC &
geometry\tablenotemark{a} & 
$n_0$\tablenotemark{b} &
$T_0$\tablenotemark{c} &
$\xi_n$\tablenotemark{d} &
$\sigma_{res}$\tablenotemark{e} &
$\Delta$\tablenotemark{f} \\
name & & (cm$^{-3}$) & (K) & & (km s$^{-1}$) & \\
\hline
%\begin{deluxetable}{lrccccc}
%\tablecaption{Summary of fit results }
%\tablewidth{0pt}
%\tablehead{
%\colhead{IRDC} & 
%\colhead{geometry$^a$} & 
%\colhead{$n_0^b$} & 
%\colhead{$T_0^c$} & 
%\colhead{$\xi_n^d$} & 
%\colhead{$\sigma_{res}^e$} & 
%\colhead{$\Delta^f$} \\
%\colhead{name} & 
%\colhead{} & 
%\colhead{(cm$^{-3}$)} & 
%\colhead{(K)} & 
%\colhead{} & 
%\colhead{(km s$^{-1}$)} & 
%\colhead{}
%}
%\startdata
G005.85$-$0.23 &
sphere &
$1.1\times10^6$ &
$5.0\times10^3$ &
$1.1$ &
$2.4$ &
$0.18$
\\
G009.28$-$0.15a &
filament &
$4.0\times10^5$ &
$5.4\times10^3$ &
$2.3$ &
$2.5$ &
$0.23$
\\
G009.28$-$0.15b &
filament &
$8.9\times10^5$ &
$1.2\times10^4$ &
$5.5$ &
$3.8$ &
$0.27$
\\
G009.86$-$0.04 &
filament &
$6.8\times10^5$ &
$2.0\times 10^3$ &
$6.2$ &
$1.6$ &
$0.28$
\\
G023.37$-$0.29 &
sphere &
$6.0\times10^5$ &
$6.0\times10^3$ &
$1.7$ &
$2.6$ &
$0.31$
\\
G024.05$-$0.22 &
sphere &
$5.5\times10^5$ &
$6.0\times10^3$ &
$1.7$ &
$2.7$ &
$0.4$ 
\\
G034.74$-$0.12 &
filament &
$3.4\times10^5$ &
$2.3\times10^4$ &
$2.0$ &
$5.3$ &
$0.18$
\\
%\enddata
%\tablecomments{
%$^a$ Fitted geometry (BE-sphere or isothermal cylinder). \\
%$^b$ Fitted central density. \\
%$^c$ Fitted isothermal temperature. \\
%$^d$ Normalized stability parameter. $\xi_n=\xi_{ml}$ for cylinders, and $\xi_n=\xi_{BE}/6.5$ for BE-spheres. %$\xi_n>1$ indicates instability. \\
%$^e$ Residual velocity dispersion needed at minimum to support the cloud energetically.\\
%$^f$ Normalized rms error of fit. }
%\end{deluxetable}
\hline
\end{tabular}
\tablenotetext{1}{Fitting geometry (BE-sphere or isothermal cylinder).}
\tablenotetext{2}{Fitted central density.}
\tablenotetext{3}{Fitted isothermal temperature.}
\tablenotetext{4}{Normalized stability parameter. $\xi_n=\xi_{ml}$ for cylinders, and $\xi_n=\xi_{BE}/6.5$ for BE-spheres. $\xi_n>1$ indicates instability.}
\tablenotetext{5}{Residual velocity dispersion needed at minimum to support the cloud energetically.}
\tablenotetext{6}{Normalized rms error of fit.}
\end{center} 
\end{table*}

\subsection{Kinematics of ``starless'' IRDCs}

Because the feedback from embedded protostars can confuse the kinematic signatures in the IRDCs, here we take a closer look at the IRDCs which show little or no evidence of actively forming stars. Our sample includes one IRDC, G005.85$-$0.23, which lacks any coincident {\em Spitzer} point sources, thus does not appear to host embedded star formation.  G024.05$-$0.22, has one Class II object (with no 24~$\mu$m counterpart, see Figure~\ref{fig:g2405_3plot}) near the edge of the NH$_3$ (1,1) emitting region, thus the bulk of the IRDC appears devoid of YSOs.  It is possible in both cases that embedded protostars in the IRDC are heavily extincted by dust beyond our detection limit, but we continue our discussion assuming that the kinematics are dominated by the global forces rather than protostellar feedback. 

Both of these IRDCs have nearly round projected morphologies.  In G005.85$-$0.23, the gradient is from southwest to northeast (33$^{\circ}$ east of north) centered on the integrated intensity peak. In G024.05$-$0.22 the gradient is from northeast to southwest (8$^{\circ}$ east  or north) but is not symmetric about the peak in integrated intensity.  In this case, the velocity gradient is not smooth, but has a sharp ridge structure in the east-west direction, which also corresponds with enhanced linewidths ($\sim$3.6~km~s$^{-1}$ compared to $<$~2~km~s$^{-1}$ throughout the rest of the cloud). 

Smooth velocity gradients are often interpreted as signatures or rotation \citep[e.g.][]{Arquilla1986, Goodman1993}.  Since the projected geometry of both of these IRDCs is roughly circular, we can reasonably approximate the clouds as spheres.   If we assume solid-body rotation, the resulting velocity gradients are 2.4 and 2.1~km~s$^{-1}$~pc$^{-1}$ for G005.85$-$0.23 and G024.05$-$0.22, respectively.  If this organized motion is linearly fit and subtracted from the centroid velocity field, the residuals are less than 0.2~km~s$^{-1}$.  

We can then compare the importance of rotational kinetic energy to gravitational energy, parameterized by the ratio, $\beta$ (in this case for a uniform density sphere), defined as $\beta = \Omega^2 R^3 / (3 G M )$, where $\Omega$ is the angular velocity, $R$ is the cloud radius, and $M$ is the cloud mass \citep{Goodman1993}.  For the mass, we take values from \citet{Ragan_spitzer} using 8~$\mu$m absorption as a mass-tracer over the region mapped in ammonia. For reference, a value of $\beta$ = $\frac{1}{3}$ is equivalent to breakup speed for a spherical cloud, and lower values signify a lessening role of rotation in cloud energetics.  Under these assumptions, we find $\beta$ values of 2 $\times$ 10$^{-4}$ and 5 $\times$ 10$^{-4}$, lower than the extremely low end of the $\beta$ range seen in dark cores \citep{Goodman1993} because the masses are much higher (500 and 2500 $\msun$ respectively). Adopting a more realistic centrally peaked density profile, $\rho \propto r^{-2}$ for example, reduces $\beta$ by a factor of 3.  In this simplistic picture, rotation plays only a small role in the dynamics of the cloud.  

Such a simple solid-body rotation model for such spherical IRDCs is a tempting interpretation, but one that should be made cautiously.  \citet{Burkert2000} show that large-scale turbulent motions (i.e. turbulence with a steep power spectrum) can lead to centroid velocity gradients that look like shear or rotation.  Indeed, even in these IRDCs which appear to be the most quiescent, the linewidths far exceed the thermal sound speed in the typical IRDC environment (about 0.2~km~s$^{-1}$).  Such linewidths could be caused by outflows from a low-mass stellar component not detectable in the \citet{Ragan_spitzer} {\em Spitzer} observations.  Even in the absence of stellar feedback, non-thermal linewidths would be expected as consequence of the cloud formation \citep[e.g.][]{Vazquez2007,Heitsch2008a}, or if global gravitational collapse dominates the cloud evolution \citep[e.g.][]{BurkertHartmann2004,Field2008}. Below, we examine these possibilities in greater detail.

\begin{figure*}
\begin{center}
  \includegraphics[width=0.8\textwidth]{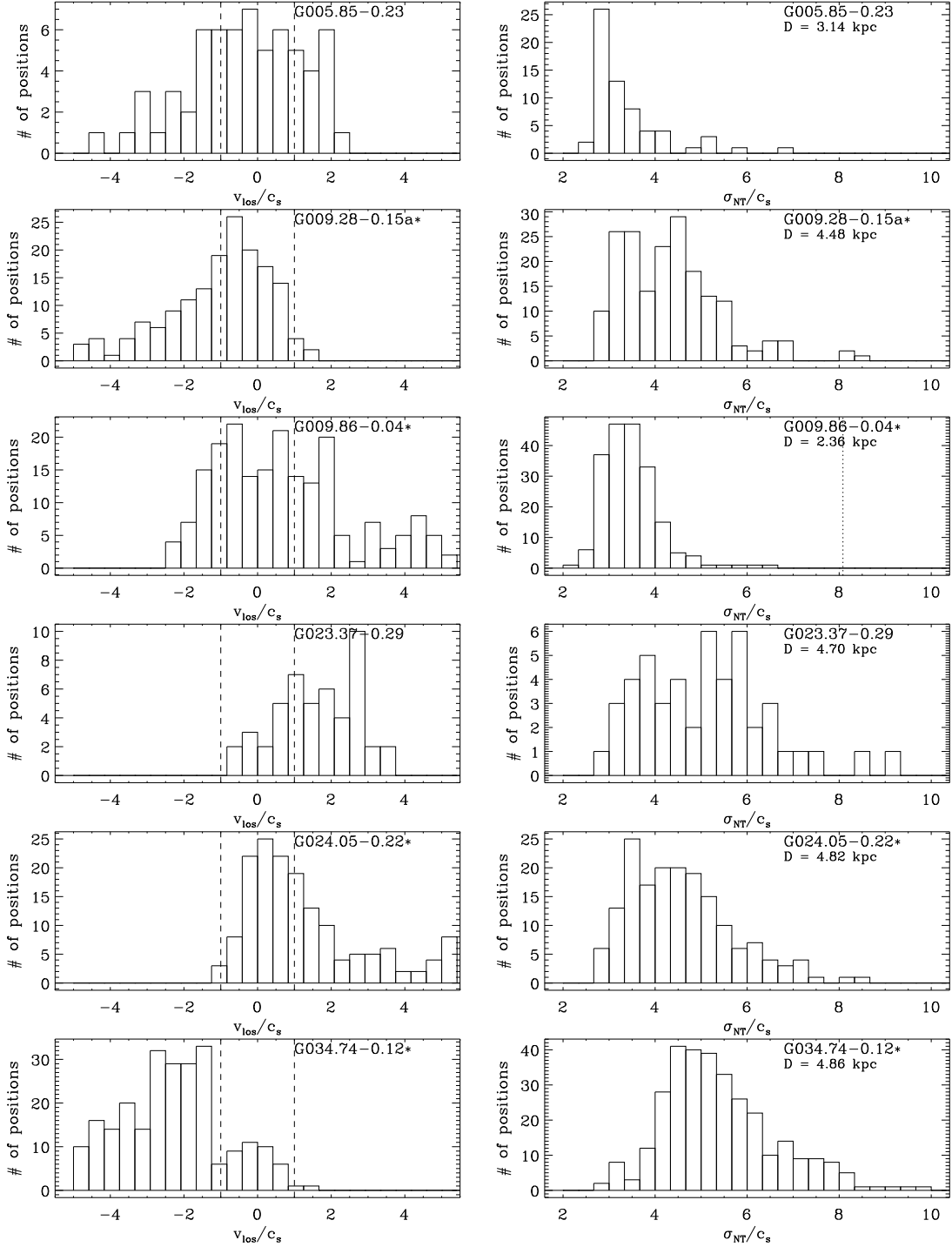}
\end{center}
\caption{\label{fig:vhist}Summary of IRDC kinematics. {\em Left column:} Histogram of the centroid velocity in units of the sound speed. Vertical dashed lines indicate the transition between sub- and super-sonic. {\em Right column:} Histograms of the non-thermal velocity dispersion (eq.~\ref{e:sigmant}). All values are supersonic by at least a factor of $2$.}
\end{figure*}

\subsection{Dynamical conditions of IRDCs} \label{ss:dynamical}

To refine our energetics estimates of the previous section, we analyzed the spatial energy distribution within the clouds. To date, dynamical studies of IRDCs have been limited mainly to single-dish surveys with resolution elements of $\sim$30$''$, which is insufficient to resolve the relevant (sub-parsec) scales. With our VLA dataset, we have mapped the velocity field across an IRDCs, so we are now able to quantify the energy distribution in IRDCs. To determine the degree of stability, we will fit (idealized) geometries, guided by our classification of the IRDCs in "spheres" and "filaments" (see Tab.~\ref{tab:fits}). The fits result in density profiles, isothermal temperatures, and a criticality parameter (see below). We use the fitted temperatures as a measure of the energy required to balance gravitational and kinetic energy content, which can be compared to the measured temperatures from Paper 1. The fit results are summarized in Table~\ref{tab:fits}.

\subsubsection{``Filaments''}\label{sss:filaments}
For the filament-like clouds, we construct radial mass density profiles from the dust column densities assuming an isothermal cylinder as the underlying model. The density profile for an isothermal cylinder is given by
\begin{equation}
  \rho(r) = \frac{\rho_0}{(1+(r/H)^2)^2},
  \label{e:isocyl}
\end{equation} 
with the cylinder scale height
\begin{equation}
  H^2\equiv\frac{2 c_s^2}{\pi G \rho_0}, 
\end{equation}
\noindent \citep[e.g.][]{Ostriker1964} with the sound speed $c_s\equiv\sqrt{k_BT_0/(\mu m_H)}$.  The fitting is done in two steps. First, we construct a filament following the mass distribution as traced by dust extinction with {\em Spitzer} \citep{Ragan_spitzer} by calculating the center-of-mass positions along the RA and DEC axes. A linear regression through the resulting positions results in the filament axis, from which the distances of the sample positions are calculated.  This gives us a radial column density profile $N_{obs}(r)$. Next, for the fitting, we start with an initial guess of $\rho_0$, $T_0$ and $R_0$, construct a profile using equation~(\ref{e:isocyl}) and project it to generate a column density profile $N_{cyl}(r)$. The rms difference
\begin{equation}
  \Delta_{rms} = \sum_i (N_{obs}(r_i)-N_{cyl}(r_i))^2
  \label{e:deltarms}
\end{equation} 
between the data and this profile is then minimized by a down-hill simplex method in the three parameters $\rho_0$, $T_0$, and $R_0$. For each cloud, we check a map of $\Delta_{rms}$ to ensure that the fit does not converge on a local minimum (it never did).

To estimate the degree of stability, we compare the fitted temperature, $T_0$, which range between 2.0 $\times 10^3$ and 2.3 $\times 10^4$~K, to the cloud temperatures presented in Paper 1 -- typically between 8 and 13~K. 
We convert the difference in temperatures to a ``residual velocity dispersion'' ($\sigma_{res}$) that would be needed to support the cloud against collapse, from 1.6 to 5.3 km s$^{-1}$. These parameters are summarized in Table~\ref{tab:fits}.
We also take the ratio of the mass per unit length derived from the fit over the corresponding critical value \citep{Ostriker1964}, 
\begin{equation}
  \xi_{ml} \equiv \frac{m}{m_c} = \frac{\pi G R_0^2\rho_0}{2 c_s^2}\frac{16H^2}{16H^2+R_0^2},
\end{equation}
with a mean molecular weight of $\mu=2.36$. Note that the fitted temperature $T_0$ is {\em not} the temperature needed to stabilize the filament against collapse, since we fit all three parameters, central density, temperature, and radius. Thus, for $\xi_{ml}>1$, the filament will collapse.

Figure~\ref{fig:isocyl} summarizes the results. The left column shows the dust column profiles (symbols) and the best fit (solid line) with the corresponding parameters. As is already suggested by the maps, there is a substantial scatter in the profiles, resulting in relative rms errors between $20$ and $30$\%.  Note that we split up G009.28$-$0.15 into two components divided at $\delta=-21^\circ$~$00'$~$20"$, just as was done in Paper 1.  The non-thermal velocity dispersion profiles (Equation~\ref{e:sigmant}) are given in the right column, with the data shown in symbols, a linear regression in a dashed line, and a binned version in the solid line (including error bars). The dot-dashed line indicates the residual velocity dispersion that would be (at minimum) required for cloud turbulent support given the nominal fit temperature and the (observed) actual temperature.

All ``filaments" are gravitationally unstable by at least a factor of $2$ in terms of masses, consistent with the virial estimates made above. The non-thermal velocity dispersions $\sigma_{NT}$, typically between 0.5 and 2 km s$^{-1}$ (corresponding to hundreds of Kelvin), are at least a factor of $2$ below the formally required value for support, $\sigma_{res}$. If $\sigma_{NT}$ were to give rise to an isotropic pressure as envisaged in models of turbulent support \citep{mckee_tan02,mckee_tan03}, then its gradient is inconsistent with such a scenario for at least two IRDCs (G023.37$-$0.29 and G009.28$-$0.15b). The remaining IRDCs show strong scatter in the velocity profiles, indicating that there is no systematic velocity distribution. 

\subsubsection{"Spheres"}
IRDCs with roughly circular shapes (or at least not obviously filamentary ones) we approximate by a (projected) Bonnor-Ebert sphere. The same reasoning as in \S\ref{sss:filaments} applies, namely that we are interested in the nominal temperature required to provide pressure support against collapse. We solve the modified Lane-Emden equation \citep{ebert,bonnor}
\begin{eqnarray}
  \frac{d}{d\xi}\left(\xi^2\frac{d\psi}{d\xi}\right)=\xi^2e^{-\psi}\\
  \xi \equiv \frac{R}{c_s}\sqrt{4\pi G n_0\mu m_H}\label{e:xi},
\end{eqnarray}
%fh: not needed
%in the standard way by splitting it into
%\begin{eqnarray}
%\frac{dy_2}{d\xi}=\frac{y_1}{\xi^2}\\
 % \frac{y_1}{\xi}=\xi^2\,e^{-y_2}
%end{eqnarray}
%(with $y_1\equiv \xi^2\eta$, $y_2\equiv \psi$, and $\eta\equiv d\psi/d\xi)$), and
integrating the ordinary differential equations with a 4th order Runge-Kutta scheme, with initial conditions $\psi(0) = 0$ and  $d\psi(0)/d\xi = 0$. A solution is determined by the central core density $n_0$, the isothermal temperature $T_0$, and the radius $R_{max}$. For $R>R_{max}$, the density is assumed to drop to irrelevant values, with the temperature increasing to provide pressure equilibrium at $R_{max}$. For values of $n_0/n(R_{max})> 14.3$, or $\xi_{BE} > 6.5$, the BE-sphere is gravitationally unstable and will collapse. 

In a procedure similar to the filament-fitting, we construct radial column density profiles from the dust data, which we then fit with a two-dimensional projection of a BE-sphere. We chose to fit all three parameters, $n_0$, $T_0$ and $R_{max}$, instead of constraining the fits by the observed cloud radius $R_{obs}$ and the central column density, because having the BE-sphere extend to $R_{max}=R_{obs}$ results in compact column density profiles, i.e. the column density would drop to zero at $R_{obs}$, inconsistent with the shape of the observed column density profiles. Thus (again), the resulting BE temperatures are {\em not} the temperatures needed to support the sphere, but they are generally smaller.  

\begin{figure*}
\begin{center}
\includegraphics[width=0.8\textwidth]{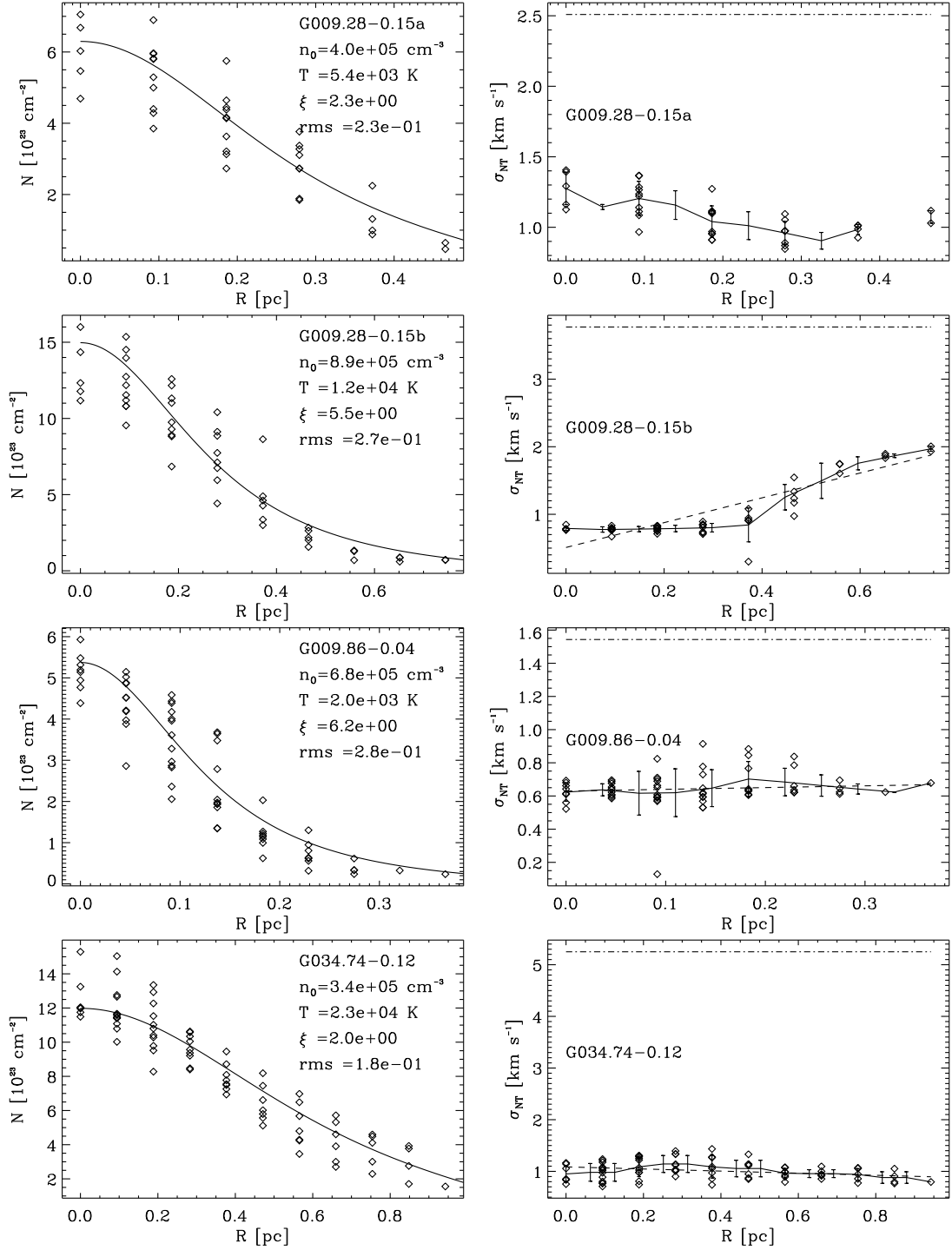}
\end{center}
\caption{\label{fig:isocyl}{\em Left:} Observed column density profiles (symbols) and best fit (line) for the three "filamentary" IRDCs, with G009.28$-$0.15 split up in a North (a) and South (b) component. Central density $n_0$, isothermal temperature $T_0$ and the ratio of mass per unit length over critical mass $\xi$ are indicated for each fit. {\em Right:} Corresponding non-thermal velocity dispersion profiles for each IRDC (symbols). A linear regression is indicated by the dashed line, and a binned profile by the solid line including error bars. The dot-dashed constant line shows the "turbulent residual velocity" necessary to stabilize the cylinder. From the $\xi$-values and the difference between actual and residual dispersion we see immediately that all IRDCs are unstable.}
\end{figure*}

\begin{figure*}
\begin{center}
\includegraphics[width=0.8\textwidth]{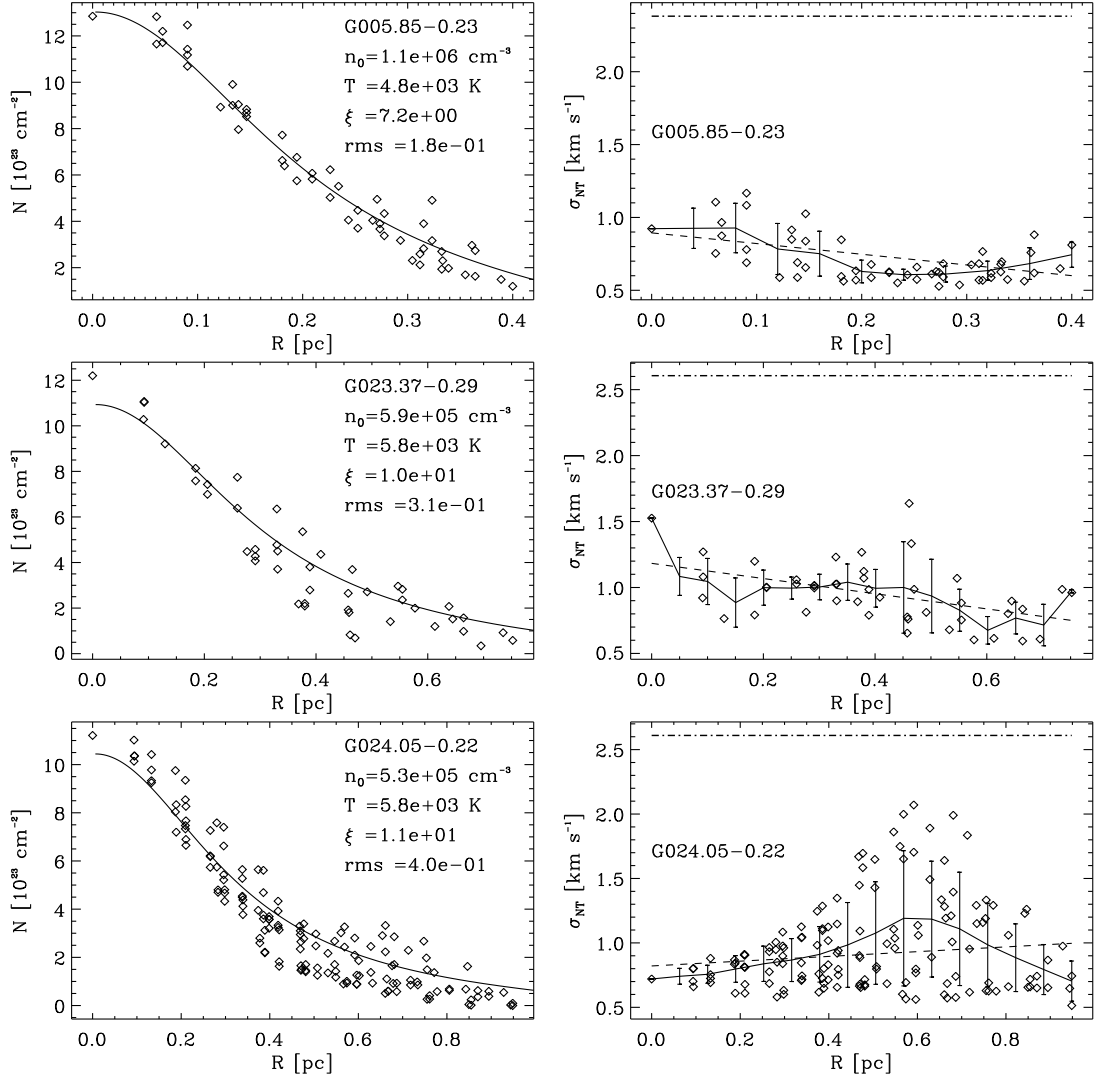}
\end{center}
\caption{\label{fig:isosph}{\em Left:} Observed column density profiles (symbols) and best fit (line) for the three "spherical" IRDCs modeled as Bonner-Ebert spheres. Central density $n_0$, isothermal temperature $T_0$, and $\xi_{BE}$ are indicated for each fit. {\em Right:} Corresponding non-thermal velocity dispersion profiles for each IRDC (symbols). A linear regression is indicated by the dashed line, and a binned profile by the solid line including error bars. The dot-dashed constant line shows the "turbulent residual velocity" necessary to stabilize the BE-sphere. From the $\xi$-values and the difference between actual and residual dispersion we see that all IRDCs are unstable.}
\end{figure*}

Figure~\ref{fig:isosph} summarizes the results analogously to Figure~\ref{fig:isocyl}. We show the radial column density profiles, the best fit and its parameters, and the velocity profiles with a linear regression, a binned profile, and the turbulent residual velocity needed to support the cloud. Comparing this to the actual dispersions gives us a measure of cloud stability. Again the fitted temperatures are on the order of 5-6 $\times 10^3$~K.

All fits result in unstable BE-spheres. The $\xi_{BE}$-values are super-critical even at the nominal (fitted) temperatures which are more than two orders of magnitude larger than the observed temperatures. The $\sigma_{NT}$-profiles are more than a factor of $2$ below the dispersion required for support, and the observed values show again a substantial scatter, indicating a strong degree of anisotropy in the spatial kinetic energy distribution. 

\begin{figure*}
  \includegraphics[width=\textwidth]{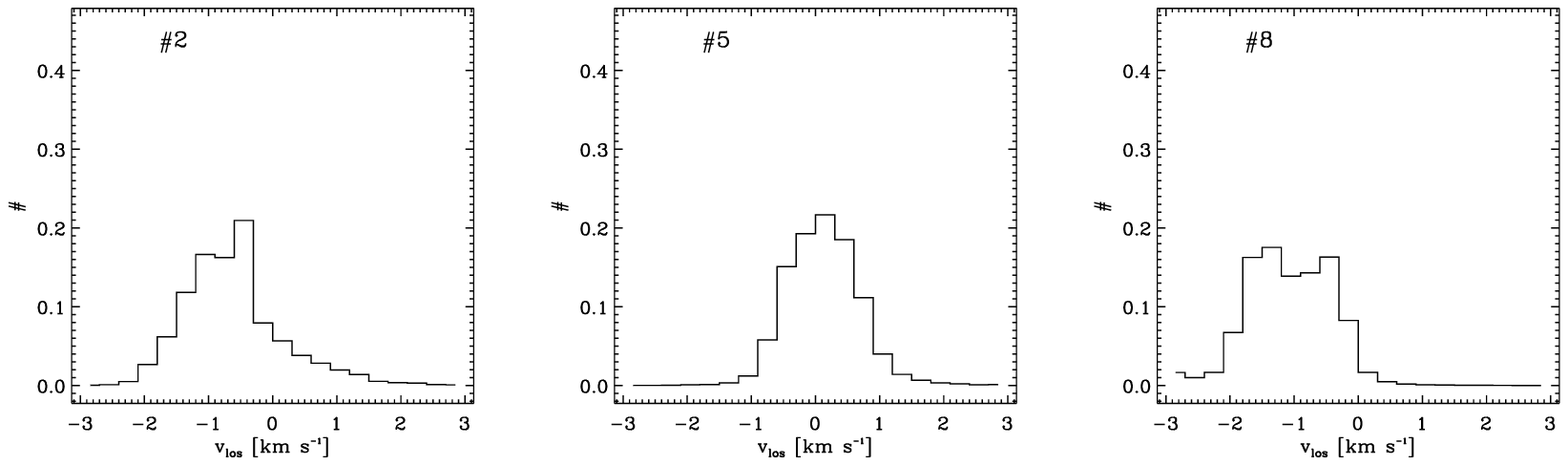}
  \caption{\label{fig:modelhisto}Normalized centroid velocity histograms of collapsing and fragmenting molecular clouds (see text). The measurements are centered on the three most massive gravitationally bound cores (numbers 2, 5 and 8 from model Gf2 of \citealp{Heitsch2008b}). Asymmetries and high-velocity tails are visible similar to the observational histograms. While not proof, the consistency of distributions is suggestive.}
\end{figure*}

\subsubsection{Implications of fitting results}

The above discussion leads us to conclude the following. All IRDCs in our sample are gravitationally unstable, with $\xi_{ml}>1$, $\xi_{BE}>6.5$ (see Tab.~\ref{tab:fits}). The fitted temperatures are at least two orders of magnitude higher than the observed values for all IRDCs, yet the fits still result in unstable structures. Interpreting these temperatures as ``turbulent temperatures", the kinetic energy (including systematic motions as indicated by the centroid velocity) in the clouds is insufficient to provide support against collapse, consistent with our earlier virial estimates. The scatter of the velocity profiles is substantial (in cases more than $100$\%), suggesting that local gravitational motions dominate over ``micro-turbulent" motions. The latter have been repeatedly demonstrated to be inconsistent with observational and numerical evidence \citep[e.g.][]{BruntHeyer2002,Padoan2003,Brunt2009}.

Another way to view our results is to consider the ``turbulent residual velocity dispersion'' (dot-dashed line in Figs.~\ref{fig:isocyl} and \ref{fig:isosph}) as a measure for the {\em gravitational} energy within the IRDC. Then, its ratio to the observed dispersion values of a few is consistent with the ratios between gravitational and kinetic energies found in models of collapsing molecular clouds \citep{Vazquez2007,Heitsch2008b}, in which the non-thermal (``turbulent'') line widths in molecular clouds are driven by global gravitational collapse, with the kinetic energy trailing the gravitational energy by a factor of a few (see Fig.~8 of \citealp{Vazquez2007} and Fig.~10 of \citealp{Heitsch2008b}). 

Expanding on this thought, Figure~\ref{fig:modelhisto} shows centroid velocity histograms taken along three lines-of-sight from a model of flow-driven cloud formation (model Gf2 of \citealp{Heitsch2008b}). The centroid velocities were measured at a point when the cloud is gravitationally collapsing, and forming local gravitationally bound cores. The selected lines-of-sight are centered on three of the most massive cores in the Heitsch et al. study (numbers 2, 5, and 8).  The clouds form due to the collision of two warm, diffuse gas flows. Strong hydrodynamical and thermal instabilities lead to immediate fragmentation and, once a sufficiently high column density has been assembled, to local and global gravitational collapse. The model centroid profiles show similar asymmetries and tails as we see in Figure~\ref{fig:vhist}.  These asymmetries arise from infall in a non-uniform medium, i.e. clumps of gas are falling into the gravitational potential well. Such events may cause the asymmetries in the observed profiles. In other words, gravitational collapse of a molecular cloud would not necessarily result in symmetric centroid velocity distributions. Further high-resolution studies, using appropriate high-density gas tracers, must be conducted in order to test our inferences about gravitational collapse in more detail.

If turbulence cannot support the IRDCs, could magnetic fields? In the absence of observational data for our IRDCs, we can estimate the critical field strengths required for cloud support. Using the expression for the critical mass-to-flux ratio for a sheet-like cloud by \citet{NakanoNakamura}, the critical field strength is given by 
\begin{equation}
  B_{cr} = 0.34\,\left(\frac{M}{\mbox{M}_\odot}\right)\left(\frac{A}{\mbox{pc}^2}\right)^{-1}\,\,\mu\mbox{G},
  \label{e:bcrit}
\end{equation}
with the cloud mass $M$, and the (projected) area $A$. The estimates (Table~\ref{tab:targets}) are larger by a factor of a few than magnetic field strengths from CN Zeeman measurements \citep[e.g.][]{Falgarone2008,Crutcher2010}, suggesting that magnetic fields are unlikely to provide wholesale support to the IRDCs, although they might be strong enough to affect the gas dynamics.

\section{Conclusion}
In this paper, we have furthered our analysis of the ammonia maps presented in Paper 1, focusing here on IRDC kinematics. 
Our main conclusions are as follows: 

\begin{itemize}

\item
In general, the imaged kinematic properties derived from the NH$_3$ (1,1)
line and (2,2) line are very similar and strongly corroborate each other.
A notable exception is the IRDC G009.86$-$0.04 where the line center
velocities are offset by $\sim0.5$~km~s$^{-1}$ and the (1,1) linewidths
are everywhere below 2 km~s$^{-1}$ while the (2,2) linewidths reach up to
4~km s$^{-1}$; these differences are likely due to the active cluster
formation underway in this IRDC, which selectively affects slightly warmer
gas traced by the higher excitation (2,2) line.

\item
For all of the IRDCs with robust measurements, non-thermal motions are greater than
thermal motions by factors of 2 to 8. The linewidths are always greater
than the range of centroid velocities across the cloud. Indeed, objects of
this mass are expected to be in early phases of fragmentation from turbulent
molecular cloud complexes, and this phase of fragmentation is integral to
setting the conditions of the massive star and cluster formation to follow.

\item
The velocity fields across the IRDCs are typically very regular, showing
smooth gradients in centroid velocity at the resolved size scales.
These gradients could be due to rotation, shear, infall,
or residual turbulent motions from the fragmentation process. 
Observed departures from the regular
trends are generally connected to mid-infrared point sources tracing
embedded young stellar objects. At the sites of these sources, the centroid
velocity may be shifted by 0.5 to 1.5 km~s$^{-1}$, perhaps due to infall
onto, or outflow feedback from, protostars within the clouds. These effects
tend to be greatest when a point source is detected at 24 $\mu$m only, i.e.
at an early phase of star formation.

\item
For all of the IRDCs, the kinetic energy estimated from the observations
is insufficient to provide support against collapse. We perform basic
models taking into account the projected geometry of the IRDC. This spatial analysis
of the thermal, kinetic and gravitational energy content indicates that
none of the clouds are in equilibrium.
Rather, the energetics combined with the density structure suggest that the clouds
are in active fragmentation and collapse, in contrast to the static ``turbulent core'' picture outlined by \citet{mckee_tan02,mckee_tan03}.

\end{itemize}

\acknowledgements
The authors thank the anonymous referee for comments which greatly improved the manuscript. 
SR is grateful to Lee Hartmann and Fred Adams for useful discussions.
This work was supported by the National Science Foundation under Grant 0707777. 
FH acknowledges support by NSF grant AST-0807305.

\end{document}